\documentclass[11pt]{article}  
\usepackage{graphicx}
\usepackage{natbib}
\usepackage[a4paper,left=2.5cm,right=2.5cm, bottom=2.5cm, top=2.5cm]{geometry}
\usepackage{amssymb}
\usepackage{bm}
\usepackage[margin=1pt,font=small,labelfont=bf]{caption}
\usepackage{multicol}
\usepackage{subfigure}
\usepackage{lastpage}
\usepackage{subfiles}
\usepackage[usenames,dvipsnames]{xcolor}
\usepackage[unicode,hyperfootnotes=false,breaklinks=true,colorlinks=true, allcolors=blue]{hyperref}
\usepackage{mathtools}
\usepackage{setspace}
\usepackage{fancyhdr}
\usepackage{titlesec}
\usepackage{times}

\fancypagestyle{custom}
{      
\fancyhead[L]{}
\fancyhead[R]{}
\fancyhead[C]{}
\fancyfoot[C]{\thepage}
\fancyfoot[L]{}
\fancyfoot[R]{}
}

\renewcommand\thesection{\arabic{section}}
\titlespacing\section{0pt}{12pt plus 4pt minus 2pt}{0pt plus 2pt minus 2pt}

\DeclareMathOperator\arcsinh{arcsinh}
\begin{document}

\pagestyle{custom}
\begin{center}
\LARGE {\bf FASTDASH:  An Implementation of 3D Earthquake Cycle Simulation on  Complex Fault Systems Using the Boundary Element Method Accelerated by H-matrices\\[12pt]}
\large
Jinhui Cheng$^{1,*}$, Harsha S. Bhat$^1$, Michelle Almakari$^1$, Brice Lecampion$^2$, Carlo Peruzzo$^2$ \\[12pt]

\begin{enumerate}
\small
\setlength\itemsep{0.01em}
\item Laboratoire de Géologie, Ecole Normale Superieure, CNRS-UMR 8538, PSL Research University, Paris, France.\\
\item Geo-Energy Lab - Gaznat Chair on Geo-Energy, Swiss Federal Institute of Technology in Lausanne, \\EPFL-ENAC-IIC-GEL, Station 18, Lausanne CH-1015, Switzerland\\
\item Currently at Division of Geological and Planetary Sciences, California Institute of Technology, Pasadena, USA.
\end{enumerate}
\vspace{-12pt}
\textcolor{gray}{\bf Manuscript accepted with minor revisions in Geophys. J. Int.}
\end{center}

%%%%%%%%%%%%%%%%%%%%%%%%%%%%%%%%%%%%%%%%%%%%%%%%%%%%%%%%%%%%%%%%%%%%%%%%%%%

\section*{Abstract}
Fault systems have geometrically complex structures in nature, such as stepovers, bends, branches, and roughness. Many geological and geophysical studies have shown that the geometrical complexity of fault systems in nature decisively influences the initiation, arrest, and recurrence of seismic and aseismic events. However, a vast majority of models of slip dynamics are conducted on planar faults due to algorithmic limitations. We develop a 3D quasi-dynamic slip dynamics model to overcome this restriction. The calculation of the elastic response due to slip is a matrix-vector multiplication in boundary element method, which can be accelerated by using Hierarchical Matrices. The computational complexity is reduced from $O(N^2)$ to $O(N\log N)$, where $N$ is the number of degrees of freedom used. We validate our code with a static crack analytical solution and the SEAS benchmark/validation exercise from Southern California Earthquake Center. We further employ this method on a realistic fault system with complex geometry that was reactivated during the 2023 Kahramanmaraş – Türkiye doublet earthquakes, generating slip sequences that closely match real observations.

\section{Introduction}

Seismic hazard assessment is essential for public safety and infrastructure resilience, aiming to quantify seismic risks for informed decision-making and preparedness. The maximum magnitude of earthquakes, influenced by the complexity of fault systems, is a critical aspect of this assessment. The geometrical complexity of fault systems in the natural environment decisively influences the initiation, propagation, and arrest of seismic events \citep{king1985,nakata1998,wesnousky2006}. Nonplanar faults, with their multi-scale roughness, introduce stress heterogeneity that determines earthquake size by controlling rupture termination. Rupture jumps across fault stepovers can significantly increase the size of earthquakes, heightening the risk and potential damage.

Geological and geophysical investigations have shed light on the interconnected relationship between fault geometry and seismic behavior. The relationship between local geometrical discontinuities (e.g., bends and steps) in fault traces and rupture arresting has been studied by \citet{Biasi2016, Biasi2021} through analysis of mapped surface rupture traces from numerous historical earthquakes. Concurrently, mechanical analyses, emphasizing elastic interaction, underscore the pivotal role of geometry in fault mechanics \citep{Segall1980}. In addition, observations from seismic events challenge the notion of isolated fault behavior. For instance, the 1992 Mw 7.3 Landers earthquake illustrated the phenomenon of multiple faults being activated simultaneously \citep{Hauksson1993, CoheeBeroza1994, Sieh1993}. Similarly, the 2001 Mw 8.1 Kokoxili earthquake showed rupture propagation across two strike-slip segments via an extensional stepover, marked by a notable delay \citep{antolik2004}. The Kunlun Fault has experienced multi-segment ruptures in over five historical earthquakes \citep{Xu2002a}. The 2009 Mw 6.3 L’Aquila event in the Central Apennines, involving the L’Aquila segment and the Campotosto listric fault, serves as a case study for understanding the role of fault geometry, with the latter exhibiting dip variations with depth where the major events nucleate \citep{Chiaraluce2012}. The 2015 Mw 7.8 Gorkha earthquake in Nepal occurred on a gently dipping section of the Main Himalayan Thrust (MHT), and the sharp change of fault geometry (kink) in the shallow part arrests the rupture \citep{Hubbard2016}. In the 2016 Mw 7.8 Kaikoura earthquake, a number of faults ruptured simultaneously \citep{Hamling2017, Cesca2017}, underscoring the complex interactions within fault networks. \citet{Lee2024} investigates the impact of fault-network geometry on surface creep rates in California. The study finds that simpler fault geometries correlate with smooth fault creeping, whereas more complex geometries are prone to locking and exhibit stick-slip behavior, leading to earthquakes. Traditional models, which rely on laboratory-derived rate-and-state frictional parameters, often overlook the role of fault-system geometry. These examples highlight the primary role of fault geometry in governing the nucleation, propagation, and termination of earthquakes.

Therefore, to more accurately estimate earthquake magnitudes and unravel the complexities of the rupture process, it is imperative to develop refined fault models that accurately capture the subtleties of fault geometry. The realistic fault system's geometry can be comprehensively reconstructed from a variety of techniques. Field surveys provide direct observations of surface ruptures, while satellite imagery offers a broad view of affected areas. Interferometric Synthetic Aperture Radar (InSAR) offers precise surface geometry mapping through the detection of surface displacement discontinuities. Advanced technologies like 3D terrestrial laser scanning (TLS) and drones give us detailed fault geometries \citep{Wilkinson2010, LiuZeng2022}. Fault depth information, which is difficult to measure from the surface, can be inferred from the distribution of seismicity with high resolution \citep{Ross2020, Ross2022}. Other approaches like Global Positioning System (GPS) and InSAR inversion, seismic reflection profiles, and geological cross sections can also help to constrain fault depth. A combined approach has been successfully used in places such as the Central Apennines \citep{FaureWalker2021}, Turkey \citep{emre2018active}, Southern California \citep{Plesch2007}, and New Zealand \citep{Seebeck2023}.

Numerical modelling emerges as a powerful tool to study the complexities of slip dynamics within fault systems. Different from statistical models of seismicity like the Epidemic Type Aftershock Sequence (ETAS) to represent earthquake occurrences \citep{Ogata1988}, physics-based numerical models offer a comprehensive understanding of earthquake processes by solving the evolution of slip on pre-existing fault systems governed by friction laws. Precise definition of fault geometry is crucial for accurate modeling. Dynamic rupture modeling, with its accurate fully dynamic solution, primarily focuses on the coseismic process for a single event \citep{aochi2002a, duan2012, ando2017, Taufiqurrahman2023}, where the fault is governed by slip weakening friction. However, it doesn't cover the entire earthquake cycle or simulate both seismic and aseismic sequences. Seismic cycle simulations are capable of modeling the entire earthquake cycle, encompassing interseismic, preslip, coseismic, and afterslip phases over large time scales. The quasi-dynamic approach is widely used in seismic cycle simulations to reduce computational costs by replacing inertial wave-mediated effects with a radiation damping term \citep{rice1993}. The Boundary Element Method (BEM) efficiently models arbitrary fault geometries by only discretizing fault planes and assuming homogeneous materials, offering fast computations \citep[e.g.][]{Yu2018, Thompson2019}. In contrast, volume-based methods like the Finite Difference Method \citep{LiMeng2022}, Finite Element Method \citep{Liu2019}, and Discontinuous Galerkin Method \citep{Uphoff2022} have their advantages by accounting for elastic heterogeneity and inelastic bulk deformation.

Geological faults, spanning a vast range of lengths, pose a computational challenge for seismic cycle simulations. Simulating the earthquake cycle in 3D fault networks, by embedding 2D faults within a 3D medium, incurs significantly higher computational costs than 2D simulations. It is a challenge for computing. However, observational studies, such as those by \citep{Ross2020}, have revealed that earthquake swarm behavior is profoundly influenced by the three-dimensional fault structure, a detail overlooked by 2D models. This highlights the urgent need for an efficient tool capable of simulating 3D earthquake cycles within complex fault geometries to accurately comprehend slip sequences.

3D earthquake modeling often involves a vast number of variables, sometimes reaching up to a million degrees of freedom. While methods like the spectral boundary integral element method using Fast Fourier Transform (FFT) offer efficiency, they are typically limited to single planar faults \citep{lapusta2009a}. Recent advancements, such as the work by \citet{Romanetozawa2021}, aim to address mildly nonplanar faults, but these methods have not yet been extended to full 3D. \citet{Barbot2021} explored the dynamics of interaction on multi-parallel faults by solving corresponding elastostatic Green’s functions. The Fast Multipole Method and Hierarchical Matrices provide alternatives for efficient matrix-vector multiplication. The Fast Multipole Method, as elucidated by \citet{Greengard1987} and implemented by \citet{Thompson2019} in earthquake cycle simulation, serves as a powerful tool for approximating far-field elements within a dense matrix, which is constructed from discretized elastic kernels. However, its application necessitates the computation of multipole expansions for various kernel types, introducing a layer of complexity. In contrast, Hierarchical Matrices present an efficient strategy for compressing dense matrices in a purely algebraic manner \citep{Hackbusch2015}. By selectively disregarding specific far-field elements based on predefined conditions, these matrices accelerate matrix-vector multiplication in earthquake cycle modeling \citep{Ohtani2011, Luo2018, Ozawa2022}. The integration of Hierarchical Matrices in quasi-dynamic simulations offers a substantial reduction in computational complexities, scaling down from $N^2$ to a more manageable $Nlog{N}$, where $N$ is the number of elements.

This paper presents FASTDASH (FAult SysTem Dynamics: Accelerated Solver using H-mat), a 3D quasi-dynamic model designed to simulate slip sequences across complex fault systems, including both planar and non-planar faults. Our goal is to enhance the understanding of both seismic and aseismic activities in complex fault systems by integrating geological, geophysical, and mechanical perspectives. By employing Hierarchical Matrices, FASTDASH efficiently simulates complex fault networks over multiple earthquake cycles within feasible computational times. To validate FASTDASH, we benchmark it against analytical solutions for static cracks and the Sequences of Earthquakes and Aseismic Slip (SEAS) Benchmark problem BP4-QD \citep{Jiang2022}. We also demonstrate its capabilities through a simulation of the 2023 Turkey earthquake, highlighting its ability to generate complex earthquake sequences based solely on the geometrical complexities of fault systems, without incorporating rheological, frictional, or fluid interaction complexities. This approach is particularly applicable in regions where fault geometry is well-defined.

\section{Method}
\subsection{Boundary element method} \label{BEM}

For the boundary element method (BEM), only the fault planes need to be discretized and are solved under the assumptions of a homogeneous and isotropic medium. The quasi-static elastic equilibrium equation can be schematically written as
\begin{equation}
\nabla \cdot \mathbf{\sigma} = 0,
\end{equation}
where $\sigma$ is the stress tensor. The components $\sigma_{ij}$ represent stress along $x_{i}$ axis on the plane that is perpendicular to $x_{j}$. $i,j \in {1,2,3}$ for a 3D problem. Body force and inertial term are ignored here for simplification and quasi-dynamic assumption.

Hooke's law describes the material's constitutive law and relates strain and stress relations linearly as follows

\begin{equation}\label{HookesLaw}
\sigma_{ij} = C_{ijkl}\epsilon_{kl},
\end{equation}
where $C_{ijkl}$ are elastic constants, and $\epsilon_{kl}$ is the strain tensor component calculated with displacement $u$ as:

\begin{equation}\label{smallstrain}
\epsilon_{ij} = \dfrac{1}{2}\left(\dfrac{\partial u_{i}}{\partial x_{j}}+\dfrac{\partial u_{j}}{\partial x_{i}}\right).
\end{equation}

To derive the BEM formulation for elastic interactions, we need to introduce the representation theorem which is derived from the reciprocity theorem. If we know the Green's function of the medium, we can calculate the elastic displacement at any point inside the volume as follows,
 
\begin{equation}\label{representation}
u_{i}(\textbf{x}) = \int_{\Sigma} K^U_{ij}(\textbf{x}, \xi)t_{j}(\xi)d\Sigma - \int_{\Sigma} K^T_{ij}(\textbf{x}, \xi) u_{j}(\xi)d\Sigma,
\end{equation}
where \textbf{x} and $\xi$ are receiver and source point respectively; $K^U$ and $K^T$ are the displacement and the traction kernels in Kelvin's fundamental solution individually (e.g. \citet{bonnet1999} and \citet{Mogilevskaya2014}); and $u$ and $t$ represent displacement and traction fields respectively.

Consider two internal surfaces $\Sigma^{+}$ and $\Sigma^{-}$ that represent the upper and lower sides of a fault plane. Let $x_{*}$ be the distance from the fault in the fault local coordinate system. The positive side, $\Sigma^{+}$ corresponds to the side when approaching the surface from positive values of $x_{*}$. The representation theorem gives the displacement at $x_{*}^{\pm}$ boundary by taking the limit, 

\begin{equation}
u_{i}(\textbf{x}_{*}^{\pm}) = \lim_{x\to x_{*}^{\pm}}\int_{\Sigma^{^{\pm}}} K^U_{ij}(\textbf{x}, \xi)t_{j}(\xi)d\Sigma - \lim_{x\to x_{*}^{\pm}}\int_{\Sigma^{^{\pm}}} K^T_{ij}(\textbf{x}, \xi) u_{j}(\xi)d\Sigma,
\end{equation}

We assume that tractions are continuous across the fault, i.e., $t_i^+ + t_i^- = 0$. The displacement discontinuity is defined as $\Delta u = u^- - u^+$, representing the difference between the two sides of the fault, $\Sigma^-$ and $\Sigma^+$. Both the traction and the displacement discontinuity are defined on the fault plane $\Sigma$, which coincides with $\Sigma^-$ and $\Sigma^+$ in the reference configuration. Then, the displacement simplifies to:
\begin{equation}
u_{i}(\textbf{x})  = \int_{\Sigma} K^T_{ij}(\textbf{x}, \xi) \Delta u_{j}(\xi)d\xi.
\end{equation}

\begin{figure}
\centering
\includegraphics[width=\textwidth]{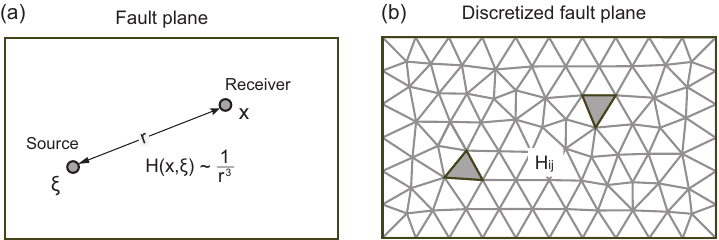}
\caption{Schematic diagram of (a) boundary integral equation for traction-slip relations (b) discretized form for traction-slip relations.}\label{fig:BEM}
\end{figure}

Substituting the above into the small strain constitutive relationship  (equations \ref{HookesLaw} and \ref{smallstrain}) the traction integral becomes

\begin{equation}
T_{i}(\textbf{x})= \int_{\Sigma} H_{ij}(\textbf{x}, \xi) \Delta u_{j}(\xi)d\xi,
\end{equation}
where traction $\textbf{T} = (T_{1},T_{2},T_{3})$, and $H$ is one of the hypersingular kernels in Kelvin solution \citep{bonnet1999, Mogilevskaya2014}. See it in Fig.  \ref{fig:BEM} (a).

The numerical implementation requires discretising the 'average' fault plane $\Sigma$ into a set of elements in which the traction-displacement discontinuities relation can be solved analytically. Each element is assumed to be either of triangular or rectangular shape, with constant displacements discontinuities $\Delta u_{j}$ over the entire element. The traction are computed analytically at the element's centroid.

In the final system of equations, the traction at one element $T_{i}$ accounts for the displacement discontinuity $\Delta u_{j}^{q}$ occurring at all elements:
\begin{equation}\label{TractionGeneral}
T_{i}(\textbf{x})= \sum_{q = 1}^{N} \int_{\Sigma^{q}} H_{ij}(\textbf{x}, \xi)d\xi \Delta u_{j}^{q},
\end{equation}
where q is the index of each element, $u_{j}^{q}$ is the displacement in j direction for $q^{th}$ element, $\Sigma_{q}$ is the surface for $q^{th}$ element. See Fig.  \ref{fig:BEM}(b)

In Eq. \ref{TractionGeneral}, $H_{ij}$ is a hyper-singular integral over element q. The traction kernel in 3D is proportional to $r^{-3}$, where $r$ is the distance between source and receiver \citep{hills2013}. The singularity comes out when the source and receiver points are overlapped. Numerical calculation of the integral is unstable in the vicinity of singularities. There are different techniques to remove the singularity. For a 2D kernel, \citet{Tada1997} integral by parts and address by Cauchy principal values. For a 3D problem, to solve elastic fields induced by a triangular dislocation loop (which is discontinuity displacement method), we use explicit formulae for 3D hyper-singular integral equation of elastostatics from \citet{Fata2011}. For \citet{Fata2011}, the surface integral is simplified to contour integral with Stokes theorem, and the hypersingular term is reduced to weakly singular term that can be evaluated in terms of Cauchy principal values. The explicit formula is written in recursive format with regard to each edge of the triangles, which is easy for numerical implementation. 

Finally the elastic traction in discretized form has a matrix vector format:  

\begin{equation}\label{TractionFinal}
T_{i} = \sum_{j = 1}^{N} A_{ij}\Delta u_{j},
\end{equation}
where $T_{i}$ is the $i$ component of traction, contributed by each components of $\Delta u$ individually. $j$ is the index of the element. $A$ is densely populated matrix of size $N\times N$. Each entry of that matrix is an integral equation over the corresponding element.

\subsection{Governing equation}\label{governing eq}

Numerical modelling of earthquake cycle for BEM is governed by the
force equilibrium equation on the discretized fault plane. On each element, the total traction on the fault is balanced with force resistance. We use quasi-dynamic BEM, the inertial term is removed and approximated by a radiation damping term \citep{rice1993}:
\begin{align}
\boldsymbol{\tau^{rad}} = -\dfrac{\mu}{2C_s} \boldsymbol{V},
\end{align}
where $\boldsymbol{V}$ is $(V_{s}, V_{d})$, slip rate in shear and dip directions. In our model, only slip along the fault plane is allowed, with no opening (tensile) slip permitted. $\mu$ is shear modulus and $C_{s}$ is shear wave velocity.

Total traction includes three ingredients, elastic shear traction $\tau^{el}$, far field loading $\tau^{load}$ and radiation damping $\tau^{rad}$. For elastic shear traction, it can be recast to a matrix-vector format in BEM (as discussed in section \ref{BEM}). For far field loading, we allow constant plate rate loading (by using backslip approach) or stress rate loading in the same principal directions with background stress. For radiation damping, it only exists in the plane constructed by fault plane.

The force resistance $\tau^{f}$ is normal traction $T_{n}$ (compression is negative and extension is positive) times friction coefficient $f$:
\begin{equation}
\tau^{f} = -T_{n}f
\label{tf}
\end{equation}

We assume the fault is governed by rate and state friction (RSF) law \citep{Dieterich1979,Ruina1983}, and regularized form of friction is \citep{Lapusta2000,rice2001a}
\begin{equation}
f = a \cdot \mathrm{sinh}^{-1}{\left[\dfrac{V}{2V_{ref}}\exp{\left\{\dfrac{f_0+b\log{(V_{ref}\theta/Dc)}}{a}\right\}}\right]},
\label{friction}
\end{equation}
$a$ is the direct effect parameter that governs the instantaneous change in friction with a change in slip rate, $b$ is the the evolution effect parameter which controls how friction evolves over time via changes in the state variable, $D_{c}$ is the characteristic slip for state evolution, $V_{ref}$ is reference slip rate and $f_{0}$ is the reference friction at $V_{ref}$. 

With ageing law to describe evolution of state variables $\theta$:
\begin{equation}
\dfrac{d \theta}{d t} = 1-\dfrac{V\theta}{D_c}.
\label{theta}
\end{equation}

Rake angle $\lambda$ is defined as angle between slip vector and strike vector. We assume that the slip vector is parallel with the traction vector and we allow rake angle rotation. The force equilibrium equation in 3D can be written as:

\begin{equation}
\tau^{f}\cos{\lambda} = \tau_{s}^{el}+\tau_{s}^{load}+\tau_{s}^{rad} \qquad \text{for strike direction }  \vec{s}
\label{t1}
\end{equation}

\begin{equation}
\tau^{f}\sin{\lambda} = \tau_{d}^{el}+\tau_{d}^{load}+\tau_{d}^{rad} \qquad \text{for dip direction }  \vec{d}
\label{t2}
\end{equation}

\begin{equation}
T_{n} = \tau_{n}^{el}+\tau_{n}^{load}
\qquad \text{for fault normal direction }  \vec{n}.
\label{t3}
\end{equation}

By coupling with rate and state friction law and state variable evolution from Eq. \ref{friction} and Eq. \ref{theta}, we have five equations to describe the physical variables changes on the fault with time in order to ensure satisfaction of the force equilibrium. We differentiate each of them with time and get a set of explicit ordinary differential equations (ODEs).

\begin{equation}
\left\{\dfrac{d\textbf{y}}{dt}\right\} = \left[\mathcal{M}\right]\left\{\textbf{y}\right\},
\end{equation}
where $\textbf{y} = (V,\lambda, T_{n}, \tau, \theta)$, and $\mathcal{M} \in \mathbb{R}^{5\times5}$. Please refer to Appendix \ref{ode} for more details on the derivation of the ODE system.

There are 5 unknowns in this system including slip rate, rake angle, normal traction, shear traction and state variable. We solve this system for every element using Runge-Kutta (RK45) method \citep{Fehlberg1969,Press2007}. RK45 is an adaptive time stepping method and the time step is error-controlled, given by the relative difference between fifth-order and fourth-order Runge-Kutta solutions. For coseismic phases, time steps can be very small in order to capture the rapidly changing slip rate and traction in the solutions.   

\subsection{Hierarchical matrices}

Introducing a third spatial dimension in geophysical simulations significantly increases computational cost. For example, simulating the fault system involved in the 2023 Turkey earthquakes, as described in Section 5, requires approximately 1,000 elements in a 2D model. In contrast, a corresponding 3D simulation demands around 100,000 elements, reflecting the steep growth in computational complexity. More critically, the dominant computational bottleneck is the elastic traction evaluation, which involves a matrix-vector multiplication in the boundary element method (BEM) (see Eq. \ref{TractionFinal}) at every time step. For dense matrices, the standard matrix-vector product scales as $O(N^2)$, making large-scale 3D simulations memory- and time-intensive.

To accelerate these computations, we exploit the fact that the off-diagonal entries of the matrix $\mathbf{A}$ in Eq. \ref{TractionFinal} decay with distance as $r^{-3}$, where $r$ is the separation between source and receiver elements. Owing to this decay, we can use Hierarchical Matrices (H-matrices) to compress and accelerate the simulation. H-matrices partition the dense matrix into sub-blocks, approximating well-separated blocks with low-rank matrices while keeping near-field blocks in full rank. The approximation quality is controlled by a user-defined accuracy.

In this work, we employ BigWham, a C++ library for vectorial boundary integral equations with hierarchical matrix compression \citep{Lecampion_BigWham_a_C_2025}. The H-matrix for a given fault geometry is constructed once before time integration. During each RK45 time step, the H-matrix is used to efficiently compute the elastic traction via fast matrix-vector multiplication.

This approach reduces both the memory requirements, compared to storing the full dense matrix and the computational complexity, reducing matrix-vector multiplication from $O(N^2)$ to approximately $O(N \log N)$.

\subsubsection{Structure of Hierarchical Matrices}
\begin{figure*}  
\includegraphics[width=\textwidth]{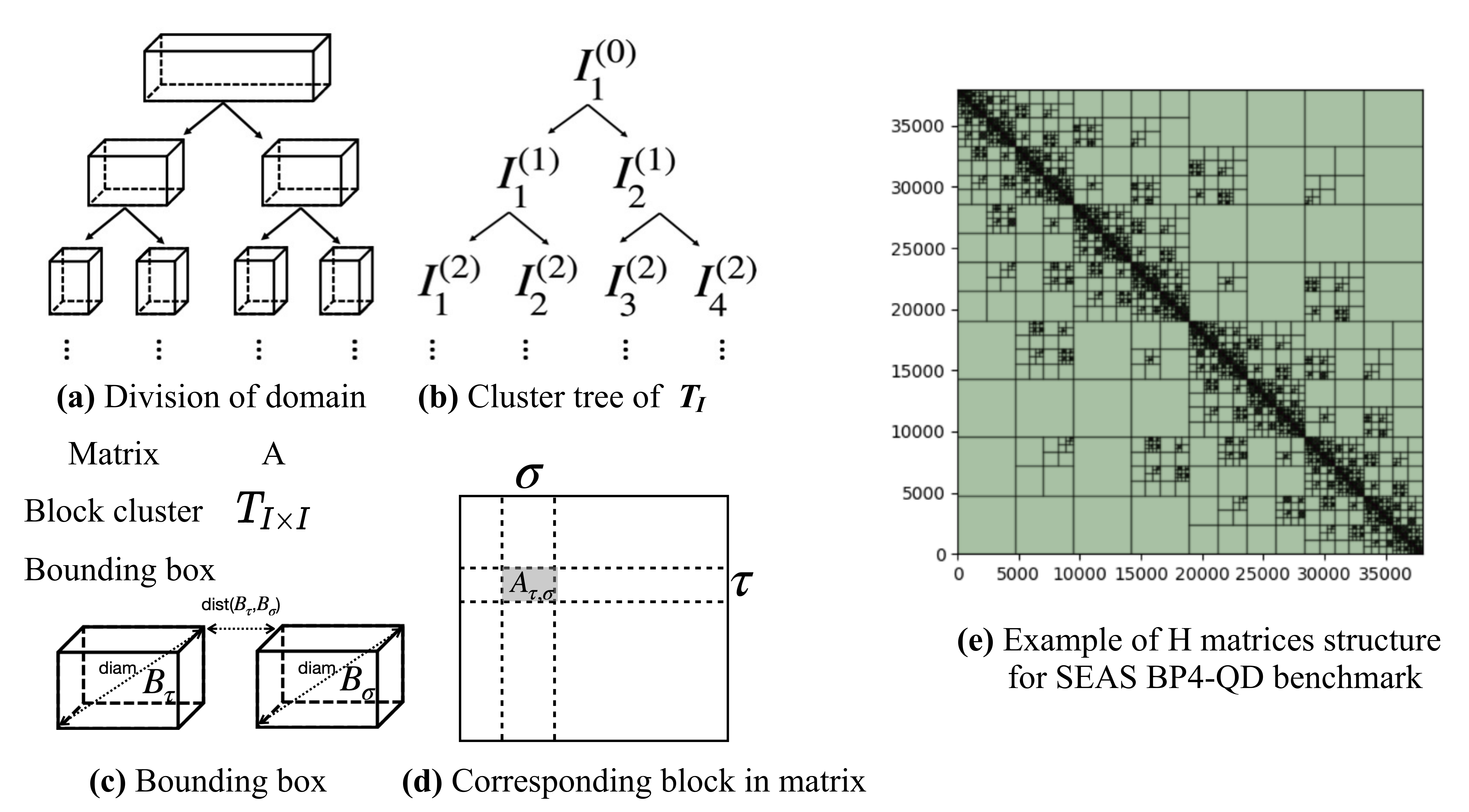}
\caption{Illustration of H-matrix structure (a) Division of domain (b) Cluster tree of $T_{I}$ (c) Initial matrix $\mathbf{A}$ can be written as a block cluster representation $T_{I\times I}$ and each cluster is enclosed into a bounding box $B$. For cluster $\tau$ and $\sigma$, the distance between the two and and each diameter are showing in the dashed lines with arrow (d) The cluster pair ($\tau$,$\sigma$) represent the row and column cluster, and define the corresponding block in matrix.}
\label{Hmat}
\end{figure*}

To compress a dense matrix using H-matrix techniques, we begin by constructing a binary \textit{cluster tree} $T_I$, based on a geometric partitioning of the mesh (Fig.~\ref{Hmat}(a)--(b)). The root cluster contains all elements in a bounding box $B$, and recursive spatial subdivision is applied via separating planes. Each node in $T_I$ corresponds to a spatial subdomain. Subdivision stops when a leaf cluster contains fewer than a user-defined threshold number of elements $N_{\text{leaves}}$.

The dense matrix $\mathbf{A}$ is then decomposed into sub-blocks, defined by pairs of row and column clusters $(\tau, \sigma)$ from the binary tree (Fig.~\ref{Hmat}(c)--(d)). Each block is tested for the \textit{admissibility condition}, which determines whether a low-rank approximation is acceptable. If admissible, compression is applied; if not, the block is further subdivided until it reaches leaf level.

Two clusters $(\tau, \sigma)$ are admissible if:
\begin{equation}
\min\left( \operatorname{diam}(B_{\tau}), \operatorname{diam}(B_{\sigma}) \right) \leq \eta \cdot \operatorname{dist}(B_{\tau}, B_{\sigma}),
\end{equation}
where $\operatorname{diam}$ is the diameter of a cluster and $\operatorname{dist}$ is the inter-cluster distance (Fig.~\ref{Hmat}(c)). The parameter $\eta$ controls admissibility strictness: larger $\eta$ values result in more aggressive compression but reduced accuracy. For elastic kernels, a value of $\eta = 3$ often yields good performance, though optimal values vary with geometry.

An example of an H-matrix structure for a 3D problem with 36,000 elements is shown in Fig.~\ref{Hmat}(e).

\subsubsection{Adaptive cross approximation}

For admissible blocks $A_{\text{LRA}}$, we can construct a low-rank approximation using Singular Value Decomposition, SVD, which provides the optimal low-rank approximation (in Frobenius norm which is the Euclidean norm of all matrix entries). The low-rank approximation is used to accelerate computation and save memory, while preserving sufficient accuracy. However, it requires full assembly of matrix entries and has computational complexity $O(\max(M,N)\min(M,N)^2)$, which is expensive. Here $M$ is the number of rows of the submatrix from cluster $\tau$ and $N$ is the number of columns of the submatrix from cluster $\sigma$.

Instead, we use \textit{Adaptive Cross Approximation} (ACA) — an iterative algorithm that constructs a quasi-optimal rank-$k$ approximation. The matrix is decomposed as:
\begin{equation}
A_{\text{LRA}} = S_k + R_k,
\end{equation}
where $S_k$ is the rank-$k$ approximation and $R_k$ is the residual. At each iteration $k$, ACA extracts a pivot row $u_k$ and column $v_k$, updating the approximation as:
\begin{equation}
S_{k+1} = S_k + u_k v_k^T.
\end{equation}

\textit{Partially pivoted ACA} requires only one row or column per iteration, making it much cheaper than fully pivoted or SVD approaches. The stopping criterion is:
\begin{equation}
\|u_{k+1}\|_2 \|v_{k+1}\|_2 \leq \epsilon_{\text{ACA}} \|S_{k+1}\|_F,
\end{equation}
where $\epsilon_{\text{ACA}}$ controls approximation accuracy and $\|S_{k+1}\|_F$ is the Frobenius norm of the matrix $S_{k+1}$ defined as $\|A\|_F = \sqrt{ \sum_{i=1}^{m} \sum_{j=1}^{n} |A_{ij}|^2 } = \sqrt{ \operatorname{trace}(A^T A) }$.

Once constructed, the low-rank block can be written as:
\begin{align}
A_{\text{LRA}} &\approx \sum_{n=1}^{k} u_n v_n^T, \\
A_{ij}^{(k)} &= \sum_{n=1}^{k} u_{ni} v_{nj}.
\end{align}

Matrix-vector multiplication with a rank-$k$ block becomes:
\begin{equation}
\sum_{j=1}^N A_{ij} \Delta u_j \approx \sum_{n=1}^{k} u_{ni} \left( \sum_{j=1}^N v_{nj} \Delta u_j \right),
\end{equation}
reducing the cost from $O(MN)$ to $O(k(M+N))$. For blocks that fail the admissibility condition, full-rank multiplication is used.

\begin{figure}
\centering
\includegraphics[width=\textwidth]{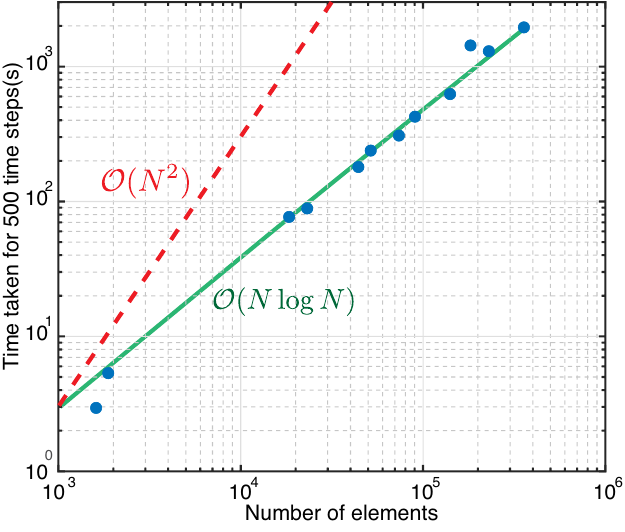}
\caption{Computational complexity. The time required for 500 time steps as a function of the number of elements. Blue dots represent our simulation results. The green line serves as a reference for \(N\log N\) scaling, while the red line represents the \(N^2\) scaling without H-matrices. N is the number of elements.}
\label{fig:computational_efficiency}
\end{figure}

In total, H-matrices reduce the overall computational complexity of matrix-vector multiplication from $O(N^2)$ to roughly $O(N \log N)$. This gain is evident in Fig.~\ref{fig:computational_efficiency}, where simulating 30,000 elements over 500 time steps drops from approximately 3000 seconds to 100 seconds. The benefit grows further with increasing model size.

 \section{Implementation}
 \begin{figure*}
\centering
\includegraphics[width=\textwidth]{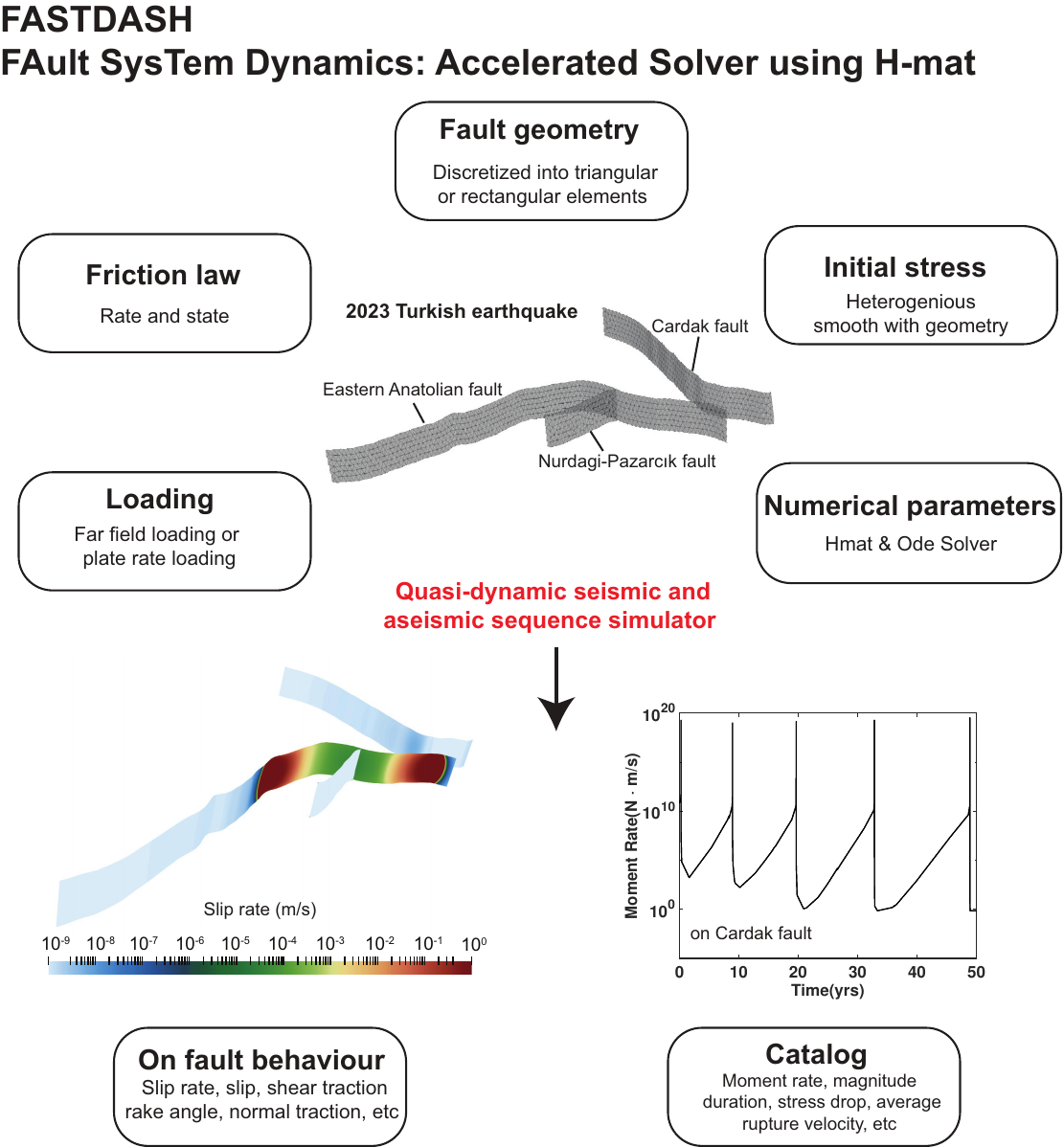}
\caption{Workflow of FASTDASH, illustrating key model input components and output results. Using the 2023 Turkish earthquake as an example, we present the real fault geometry, a slip rate snapshot, and the moment rate on the Çardak Fault for the first five earthquakes.}
\label{workflow}
\end{figure*}

FASTDASH is written in Python and can efficiently solve slip dynamics in complex fault systems. We present the workflow of this method in Fig. \ref{workflow}. Before delving into this model, we first introduce the coordinate system and sign convention in Section \ref{coord}. We discretize the natural fault system into triangular or rectangular elements as shown in Section \ref{mesh}. The fault is governed by the rate and state friction law as we discussed in Section \ref{governing eq}. Before time stepping, we assign initial conditions and loading conditions, shown in Section \ref{ICLC}. We describe how to choose the numerical parameters in Section \ref{numerical parameter}. In the end, we perform post-processing to analyze on-fault behavior (e.g., slip rate, traction, etc.) at each output time step, as well as to generate a slip catalog containing key parameters such as moment rate, magnitude, duration, stress drop, and average rupture velocity, as detailed in Section \ref{post-processing}.

\subsection{Coordinate system and sign convention}\label{coord}
\begin{figure}
\centering
\includegraphics[width=\columnwidth]{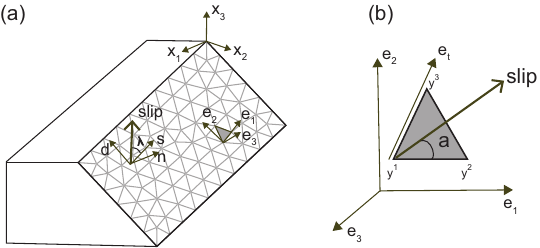}
\caption{(a) Global system $(x_1,x_2,x_3)$, local system ($e_1,e_2,e_3$) and fault system ($s,d,n$). The rake angle $\lambda$ is the angle between the strike direction $\textbf{s}$ and the slip vector (b) Local coordinate system is defined on a triangular element with three nodes $\mathbf{y}^1, \mathbf{y}^2, \mathbf{y}^3$; see Eq. \ref{local_define} for details. The rake angle in the local system $\alpha$ is defined as the angle between $e_1$ and the slip direction.}
\label{fig:coordsystem}
\end{figure}

There are three coordinate systems in this method, global system $(x_1,x_2,x_3)$, local system ($e_1,e_2,e_3$) and fault system ($s,d,n$). For global coordinate system, $(x_1,x_2,x_3)$ is a right-handed coordinate system, where the surface of the Earth is in the $x_1$ $x_2$ plane and the $x_3$ axis points vertically upwards from the Earth's surface (Fig.  \ref{fig:coordsystem}(a)). We take the real fault geometry from nature and discretize in global system with triangular (or rectangular) elements by using automatic mesh generator, CUBIT or GMSH. The mesh file with fault geometry information is written in the global system. BigWham (Hmat library) works in the local coordinate system $(e_1, e_2, e_3)$,  which is based on the triangle elements \citep{Fata2011}. There are three nodes $\textbf{y}^1, \textbf{y}^2, \textbf{y}^3$ in one triangle element, and here it requires the elements connectivity to be anti-clockwise. The orthogonal basis of local coordinate system is:
\begin{equation}\label{local_define}
\mathbf{e_1} = \dfrac{\textbf{y}^2-\textbf{y}^1}{||\textbf{y}^2-\textbf{y}^1||}, 
\vec{e_t} = \textbf{y}^3-\textbf{y}^1,
\mathbf{e_3} = \dfrac{\mathbf{e_1} \times \vec{e_t}}{||\mathbf{e_1} \times \mathbf{e_t}||}, 
\mathbf{e_2} = \mathbf{e_3} \times \mathbf{e_1},
\end{equation}
as seen in Fig.   \ref{fig:coordsystem}(b).

The fault-based coordinate system provides the most intuitive framework for analyzing the results. The unit normal to the fault plane, $\mathbf{n}$, is oriented from the negative surface to the positive surface. The strike direction, $\mathbf{s}$, is defined along the surface trace of the fault plane and is given by $\mathbf{s} = \mathbf{x_3} \times \mathbf{n}$. The third basis vector, $\mathbf{d}$, is perpendicular to both $\mathbf{s}$ and $\mathbf{n}$, pointing upward.  

The slip vector, $\mathbf{\xi}$, is defined as $\mathbf{\xi} = \mathbf{u}^{+} - \mathbf{u}^{-}$, representing the displacement of the positive surface relative to the negative surface. The rake angle, $\lambda$, is measured counterclockwise from the strike direction to the slip vector, where 0 ($\pi$) indicate left-lateral (right-lateral) motion.  

In the local coordinate system, the slip vector remains $\xi$, while the rake angle is denoted as $\alpha$, defined as the angle between $e_1$ and $\xi$.

The system of ordinary differential equations  is written in the local coordinate system of each element, such that it is easy to use BigWham to calculate the rate of elastic traction from known slip rates. To assign the initial conditions of slip rate and traction, we need to transform the coordinates from the fault-based system to the local element system. To output the slip rate and traction, we  transform back  from the local system to the fault-based system. 

The fault-based system and the local system are both element-wise and spatially varying coordinate systems, especially for nonplanar faults. For the fault-based system and the local system, $\textbf{n}$ is equal to $\textbf{e}_\textbf{3}$, which will simplify the 3D coordinate transformation into a 2D coordinate transformation. The details of the coordinate transformation are given in Eq.  \ref{transformation1}, \ref{transformation2}, \ref{transformation3} and \ref{transformation4}.

\subsection{Mesh generation}\label{mesh}

For complex fault systems, achieving high spatial resolution and effectively capturing geometrical properties can be challenging. When dealing with arbitrary shapes that require mesh generation, we employ automatic mesh generators to discretize the fault plane into triangular or rectangular elements. Among these, triangular elements offer greater adaptability to complex faults. For fundamental complexities such as stepovers and dipping faults, we recommend using GMSH \citep{Geuzaine2009}. It allows the inclusion of bounded points, lines, and surfaces with specified shapes and desired grid sizes. For realistic fault geometry, we suggest using CUBIT, which can produce a higher-quality mesh for complex fault geometries. Mesh quality control parameters (surface gradation, sizing function, condition number of mesh smoothing, and etc.) can improve the shape quality of elements, which reduces the error for benchmark exercises.
\subsection{Initial conditions and loading conditions}\label{ICLC}

We establish the pre-stress field as initial conditions. It could be given in the sense of principal stresses or background stress tensor $\sigma$. We can get traction along strike direction $T_{s}$, dip direction $T_{d}$ and normal direction $T_{n}$ by projecting onto the fault system:  
\begin{align}
\mathbf{T} = \mathbf{\sigma} \cdot \mathbf{n}\\
T_{s} = \mathbf{T} \cdot \mathbf{s}\\
T_{d} = \mathbf{T} \cdot \mathbf{d}\\
T_{n} = \mathbf{T} \cdot \mathbf{n}.
\end{align}
Because we assume that the slip vector is parallel with the traction vector,  the rake angle can be determined by the traction net vector $\mathbf{T}$ and its strike component $\mathbf{T_{s}}$.

To initiate the first event, we impose a large slip rate in a circular patch in the center of the fault to initiate the rupture nucleation. The slip rate outside of the nucleation patch is constant $V_{0}$. At steady state, $\theta = \theta_{ss} = \dfrac{D_c}{V}$. Therefore, we have the steady state initial condition of 5 unknowns [$V, \lambda, T_n,  \tau,  \theta$] for the ODE system mentioned in section \ref{governing eq}.

In FASTDASH, we have two approaches to apply the far field time dependent loading. The first approach is to consider a constant plate rate loading by using backslip method \citep{savage1983,richards2012}:

\begin{equation}
\dot{\tau} ^{load}_{i} = - V_{pl}\sum_{j=1}^{N} A_{ij},
\end{equation}
where $V_{pl}$ is plate motion velocity. Faults are embedded by creep region. The traction is calculated based on relative speed. The final equivalent changing on fault stress will be heterogeneous, increase at the edge of the fault. In the second approach we consider a constant stress loading rate, which has the same principal direction with the initial background stress tensor. Stress loading is directly projected onto the fault.

\subsection{Choice of numerical parameters}\label{numerical parameter}

The process zone is the region where breakdown energy is released, characterized by rapid variations in traction and slip. To ensure accurate results, the grid size must be fine enough to adequately resolve the process zone, enabling convergence of the solution. The process zone length, \( L_b \), is defined as:

\[
L_b = \frac{\mu D_c}{b |T_n|},
\]
where \( \mu \) is the shear modulus, \( D_c \) is the critical slip distance, \( b \) is the friction parameter describing how friction evolves, and \( T_n \) is the effective normal stress \citep{Dieterich1972}.

Grid size determines spatial accuracy, while time-step accuracy is managed using the Runge-Kutta45 method with an error tolerance. Additionally, three hierarchical matrix (Hmat) parameters are considered to ensure the accuracy of elastic traction calculations: the number of leaves (\( N _{\text{leaves}} \)), the admissibility condition parameter (\( \eta \)), and the low-rank approximation tolerance (\( \epsilon \)). A convergence test was performed using the SEAS benchmark, with details provided in Section \ref{benchmark}.

\subsection{Post-processing of on-fault data and catalog analysis}\label{post-processing}

To output full-field data, VTK files are used to store information such as slip rate and traction on the fault planes at specified intervals. These files can be visualized in 3D using ParaView, a powerful open-source visualization tool \citep{ahrens2005}.

Additionally, we can generate a slip catalog based on the on-fault data. Slow slip events and earthquakes are identified using slip rate thresholds. For slow slip events, the threshold is set one order of magnitude above the plate rate, at \(10^{-8} \, \text{m/s}\). For earthquakes, the threshold is set at \(10^{-3} \, \text{m/s}\), based on experimental and geological evidence \citep{Rowe2015}. A slow slip event is identified when the maximum slip rate exceeds the slow slip threshold but remains below the earthquake threshold. In contrast, an earthquake is identified when the maximum slip rate exceeds the earthquake threshold. The duration of the event, \(T\), is defined as the period during which the slip rate exceeds the threshold.

The stress drop $\Delta \tau$ at location $\xi$ is calculated as
\begin{equation}
\Delta \tau(\xi) = \tau^{after}(\xi) - \tau^{before}(\xi).
\end{equation}

The moment rate $\dot M$ and moment $M$ are calculated: 
\begin{align}
\dot{M}(t) = \int_{A}\mu V(\vec{\xi}, t) dA \\
M = \int_{T} \dot{M} (t) dt,
\end{align}
where $A$ is the rupture plane, $\mu$ is the shear modulus and $V$ is the slip rate.

Then the moment magnitude is calculated following \citet{Hanks1979}:
\begin{equation}
M_{w} = \dfrac{2}{3}\log_{10}(M)-6.06.
\end{equation}

\section{Benchmark/Verification}\label{benchmark}

In this section, we discuss and analyze a series of benchmark problems to verify the accuracy of the method. Static benchmark validates the elastostatic solver and dynamic benchmark validates the time stepping solver.

\subsection{Static elasto-static benchmark}

To validate our code, we first present an elasto-static benchmark which is part of the verification tests of the BigWham library. 
 To solve the elastic crack subjected to remote uniform stress loading, we superpose an uncracked body subjected to uniform far field stress and a crack loaded by internal traction.  In the case of a simple penny-shaped crack under uniform remote stress loading, the problem can be solved analytically using the Green and Collins method, which is a method to solve boundary value problem for a harmonic function. \citet{Sneddon1946} gives the analytical solution for the penny-shaped crack under tensile loading.  \citet{Segedin1951} gives the analytical solution under pure shear stress loading: 
\begin{equation}
\Delta u(r) = \dfrac{8(\lambda+2\mu)T}{\pi \mu(3\lambda+4\mu)}\sqrt{R^2-r^2}, r<R,
\end{equation}
where $\Delta u$ is the displacement discontinuity $\Delta u = u^{+}-u^{-}$. $r$ is the radius distance with the centroid. $T$ is the remote shear stress loading, $R$ is the radius of the crack. $\lambda$ is the Lamé's parameter and $\mu$ is the shear modulus. 

With FASTDASH, we can discretize the same penny-shaped crack with triangular element, apply the displacement distribution from the analytical solution and solve the traction numerically. The root mean square error reduces with increasing the number of elements. The convergence of the result is shown in Appendix \ref{penny_shaped}. 

\begin{figure*}
\centering
\includegraphics[width=\textwidth]{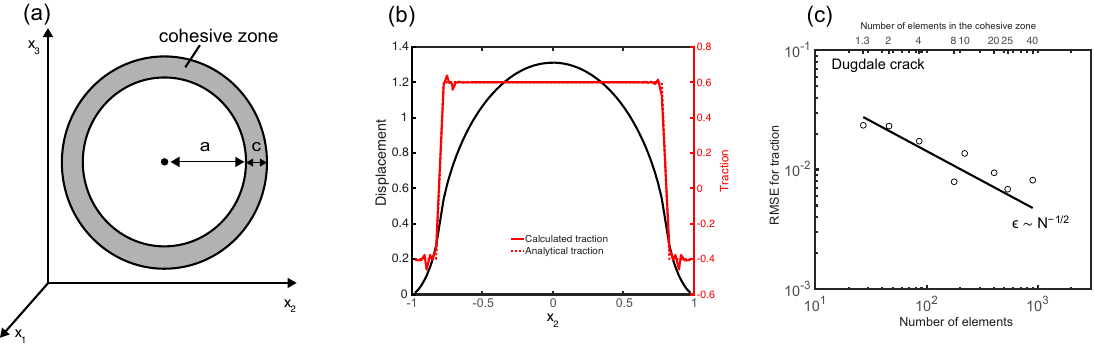}
\caption{Static crack benchmark: (a) Dugdale crack model, where an elastic penny-shaped crack with radius \(a\) is surrounded by a cohesive zone of width \(c\). (b) Slip distribution (black curve) from the analytical solution of the Dugdale crack \citep{Olesiak1968}. The red dashed line represents the analytically prescribed traction, and the red solid line shows the calculated traction. (c) Grid convergence analysis, showing the root mean square error (RMSE) for traction as a function of the number of elements along the diameter parallel to \( x_2 \). The top x-axis represents the number of elements in the cohesive zone.}
\label{fig:static_benchmark}
\end{figure*}

According to the linear elastic fracture mechanism, there are stress singularities at the edge of the crack. Therefore, we also compare our result with the Dugdale crack model, which involves an elastic circular crack surrounded by a plastic zone. Dugdale crack is loaded by a uniform remote tension $\sigma_0$. For the elastic crack $r<a$, it is traction-free but in the plastic zone $a<r<a+c$, traction is equal to the yield stress $\sigma_{Y}$. See Fig.   \ref{fig:static_benchmark}(a). The size of plastic zone $c$ is given by \citet{dugdale1960}, as follows: 

\begin{equation}
\dfrac{a}{a+c}=\sqrt{1-\dfrac{\sigma_0^2}{\sigma_Y^2}}.
\end{equation}

The displacement is given by \citet{Olesiak1968}

\small\small
\[
\Delta u(\rho) = \dfrac{4(1-\nu)l\sigma_Y}{\mu \pi}\dfrac{1}{m}\times \begin{cases}\dfrac{\sigma_0}{\sigma_Y} \sqrt{1-\rho^2}-\sqrt{\dfrac{1-m^2}{1-\rho^2}+mE(\phi_1,\dfrac{\rho}{m})} & 0<\rho<m \\
  \dfrac{\sigma_0}{\sigma_Y}  \sqrt{1-\rho^2}-\sqrt{\dfrac{1-\rho^2}{1-m^2}+\rho E(\phi_2,\dfrac{m}{\rho})}\\ 
  \qquad\qquad \qquad-\dfrac{\rho^2-m^2}{\rho}F(\phi_2,\dfrac{m}{\rho})
 & m<\rho<1
\end{cases},
\]
\normalsize
where $m = a/(a+c)$, and $\rho = r/(a+c)$. $\nu$ is the Poisson's ratio. And
\begin{equation}
\phi_{1}(\rho)=\arcsin{\sqrt{\dfrac{1-m^2}{1-\rho^{2}}}}, \phi_{2}(\rho)=\arcsin{\sqrt{\dfrac{1-\rho^2}{1-m^{2}}}},
\end{equation} 
$F(\phi,k)$ and $E(\phi,k)$ are elliptical integrals of the first kind and second kind.
\begin{equation}
F(\phi,k) = \int_{0}^{\phi} \dfrac{d\alpha}{\sqrt{1-k^{2}\sin^{2}\alpha}}, E(\phi,k) = \int_{0}^{\phi} \sqrt{1-k^{2}\sin^{2}\alpha} d\alpha.
\end{equation} 

With superposition with remote tension $\sigma_{0}$, the traction $T = \sigma_{0}$ in the elastic crack and $T = \sigma_{0}-\sigma_{Y}$ in the plastic zone. See Fig. \ref{fig:static_benchmark}(b). 

We use the analytical solution for displacement resulting from a given traction to calculate the traction and compare it with the applied traction. To ensure accuracy, we verify the root mean square error along the diameter parallel to \( x_2 \) between the numerical and analytical solutions converges as the number of elements increases, as shown in Fig. \ref{fig:static_benchmark}(c).
\subsection{Dynamic benchmark}

\begin{figure}
\centering
\includegraphics[width=\textwidth]{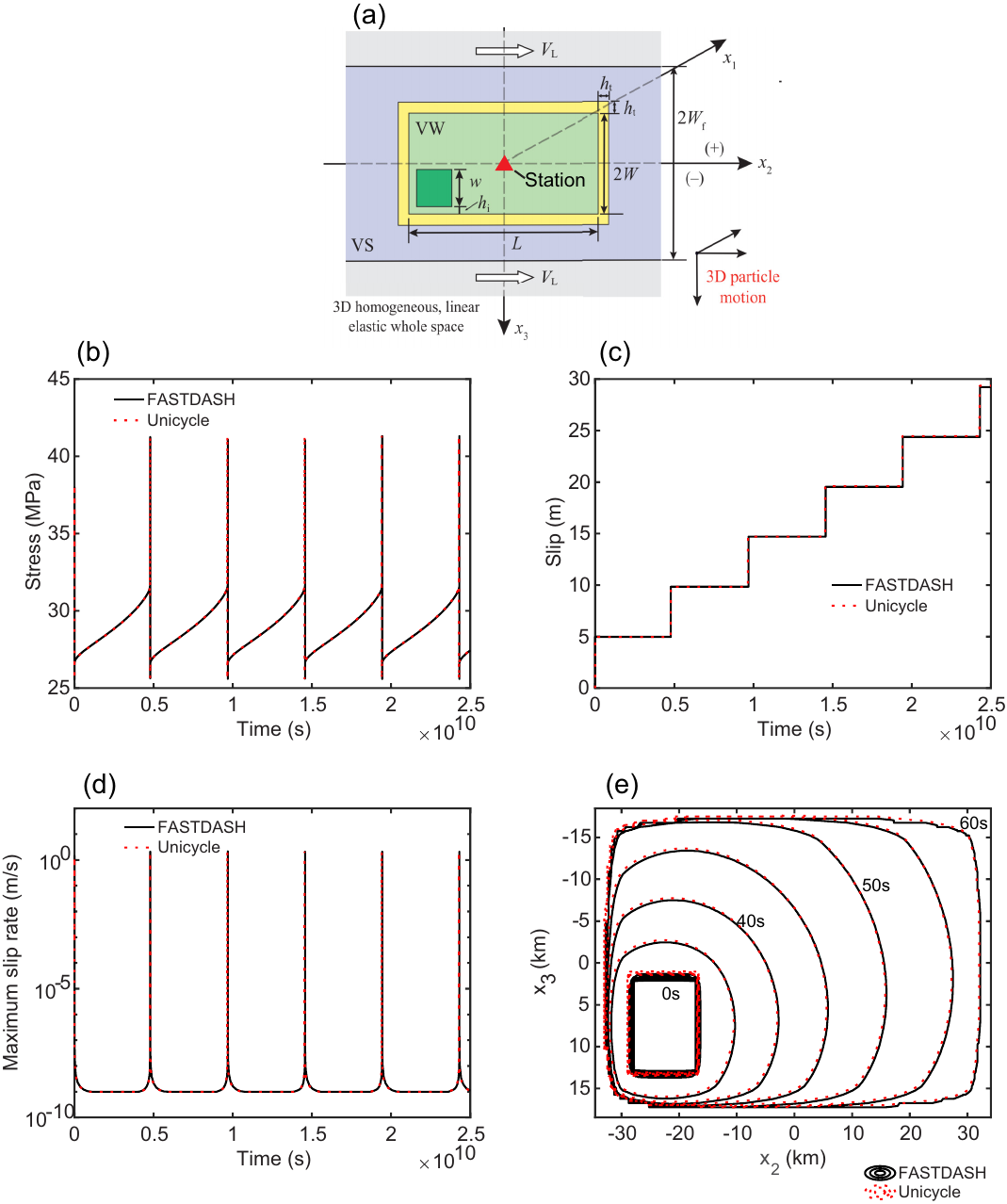}
\caption{SEAS SCEC benchmark BP4-QD. (a) Model setup \citep{Jiang2022}. (b, c) Comparison of stress and slip along strike at the station indicated in (a), between FASTDASH (black solid line) and Unicycle (red dashed line). (d) Maximum slip rate over the entire fault. (e) Rupture front contours for the first earthquake} 
\label{seas}
\end{figure}

For dynamic fault problems, there is no analytical solution. Sequences of Earthquakes and Aseismic Slip (SEAS) is a project proposed by Southern California Earthquake Center (SCEC) for validating different numerical implementation for earthquake cycles. We validate FASTDASH with SEAS Benchmark problem BP4-QD \citep{Jiang2022}. It is a problem to solve a 2D fault subjected constant plate loading rate $V_{p}$ embedded in a 3D homogeneous linear elastic whole space medium. See Fig.   \ref{seas}(a). The fault plane consists of a velocity weakening rectangle, surrounded velocity strengthening patch, with a friction transition strip in between. The initial nucleation zone (green square) is located at bottom left with a higher initial slip rate.

The use of boundary integral equations is widespread in the SEAS community. We compare our results with Unicycle, which is a boundary element method for earthquake cycle simulation without H-matrices acceleration \citep{Moore2019}. Fig.  \ref{seas} (b) and (c) shows the stress and accumulated slip on the central station. Fig. \ref{seas} (d) and (e) shows the maximum slip rate for the entire fault and rupture front contour for the first earthquake. Both local and global data shows the good agreement of the results.

We also quantify the numerical error with SEAS benchmark. We calculate the Root Mean Square Error (RMSE) to quantify the difference of rupture arrival time by using the maximum slip rate for the first six earthquakes between FASTDASH and Unicycle. The time accumulating error has already been taken into account.

\begin{equation}
RMSE = \sqrt{\sum_{i=1}^{i=6}(\Delta t_i)^2/6}/\sqrt{\sum_{i=1}^{i=6}(t_{i}^{U})^2},
\end{equation}
where $\Delta t_{i}$ is the rupture arrival time difference between FASTDASH and Unicycle for the $i$th ($i\leq6$) earthquake and $t_{i}^{U}$ is the rupture arrival time calculated from Unicycle for the $i$th ($i\leq6$) earthquake.
\begin{figure}
\centering
\includegraphics[width=\textwidth]{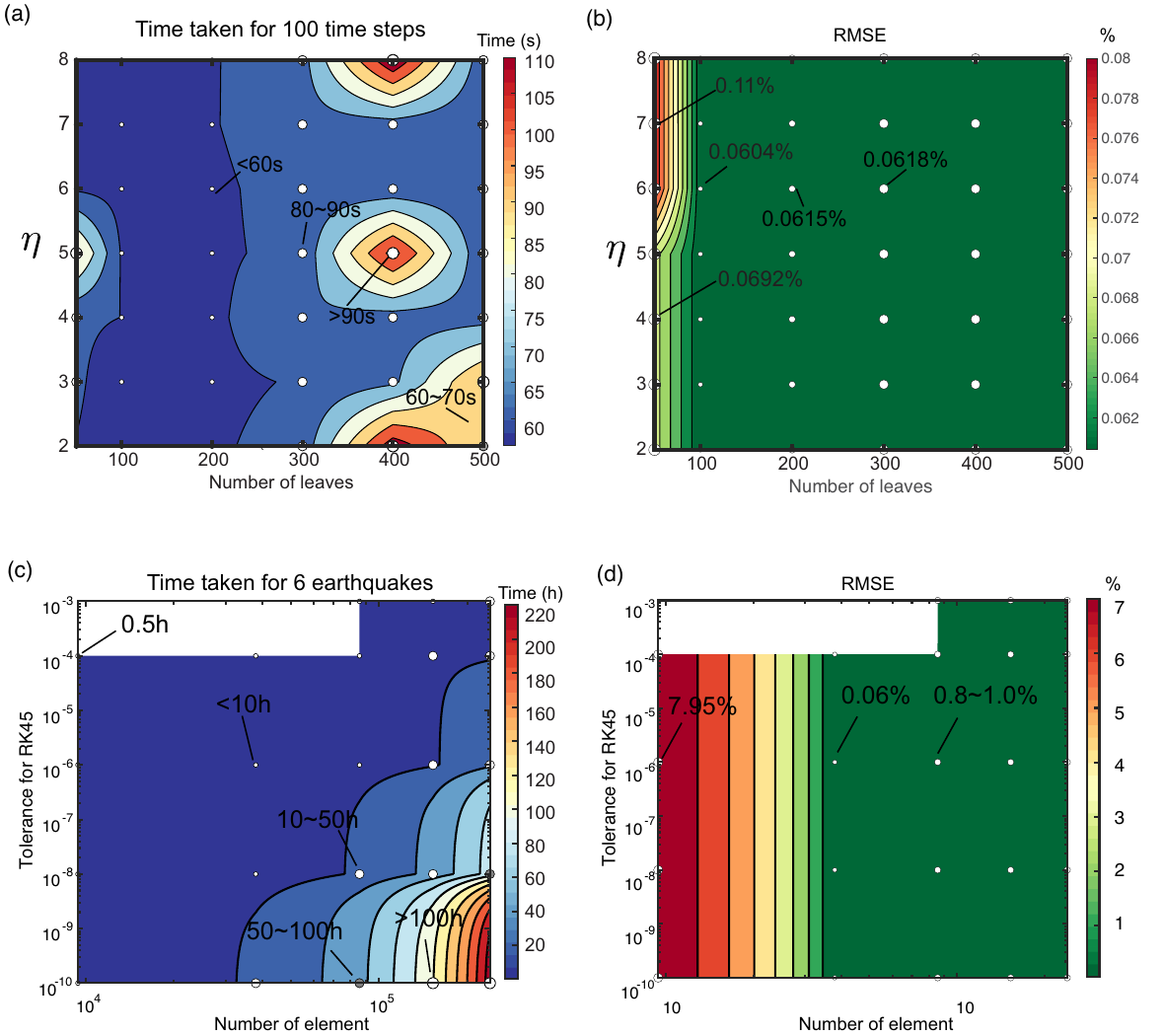}
\caption{Quantifying the numerical error with SEAS benchmark (a) The computing time for different Hmat properties (admissibility parameter $\eta$ and number of leaves $N$) (b) The Root mean square error (RMSE) relative to Unicycle for different Hmat properties (c) The computing time for six earthquakes as a function of the number of elements and tolerance of RK45. White region is where the results don't converge (d) RMSE as a function of the number of elements and tolerance of RK45}
\label{fig:SEAS_par}
\end{figure}

\begin{table}\label{tab:numerical parameters}
 \caption{Best numerical parameters for SEAS BP4-QD.}
 \centering
 \begin{tabular}{c c c}
 \hline
  Parameters & Description & Value  \\
 \hline
   $N$ & H-mat property: minimum size of blocks & 100   \\
   $\eta$ &  H-mat property: distance between accepted blocks & 3  \\
   $\epsilon$ & H-mat property: accuracy for ACA & $10^{-4}$  \\
   $\Delta s$ & Grid size &  $L_{b}/4$  \\
   $\epsilon_{rk}$ &tolerance for RK  & $10^{-4}$    \\
 \hline
% \multicolumn{2}{l}{$^{a}$Footnote text here.}
 \end{tabular}
 \end{table}

We conducted a parametric study by varying both the key parameters of the H-matrix approximation and the tolerance of the ODE solver to assess convergence (see Fig. \ref{fig:SEAS_par}). For simplicity, we set the accuracy for ACA to \(\epsilon = 10^{-4}\), as using \(\epsilon = 10^{-3}\) resulted in a significant increase in error. Theoretically, increasing the number of leaves reduces compression in the H-matrix, leading to higher accuracy but longer computation times. Conversely, a larger $\eta$ increases error because the admissibility condition allows more blocks to be approximated. 

In practice, the computation time generally increases with the number of leaves, as expected. However, for $N = 400$, the observed time increase may be attributed to the internal structure of the H-matrix. Regarding accuracy, increasing the number of leaves initially reduces error, but beyond a certain threshold (around 0.06\%), further increases in $N$ do not yield additional improvements. This suggests that the dominant source of error could be differences in time-stepping between FASTDASH and Unicycle. The results are largely insensitive to $\eta$, except for $N = 50$, where error increases with $\eta$.  

For the ODE solver, convergence is achieved at a tolerance of $10^{-4}$, beyond which errors do not further decrease. This suggests that the reference solution likely uses the same tolerance. The smallest error is observed at a grid size of 500, which closely matches the reference results.
 
The best numerical parameters for BP4-QD is listed in Table 1, where we have the most time-efficient and accurate simulations.

\section{Numerical experiment: The 2023 Kahramanmaras Turkey Earthquakes}

The February 6, 2023, earthquakes in Turkey, with magnitudes of Mw 7.8 and Mw 7.6, caused over 55,000 deaths and impacted around 14 million people, highlighting the significance of understanding fault geometry complexities for risk assessments. The earthquake of Mw 7.8 nucleated on Nurdağı-Pazarcık Fault (NPF), a branch of Eastern Anatolian Fault (EAF) and then propagated bilaterally on EAF over 300km. With a 9-hour delay, the earthquake of Mw 7.6 occurred on Çardak Fault, which varies 10 to 50km away from the EAF. All faults are left-lateral strike slip faults. In this section, we simulate earthquake cycles on the fault system that can be activated in 2023 Turkey earthquakes including NPF, EAF and Çardak fault with FASTDASH. We focus on the interaction among multiple faults and reproduce the earthquake sequences. For simplicity, we ignore other splay faults. 

\begin{figure}
\centering
\includegraphics[width=\textwidth]{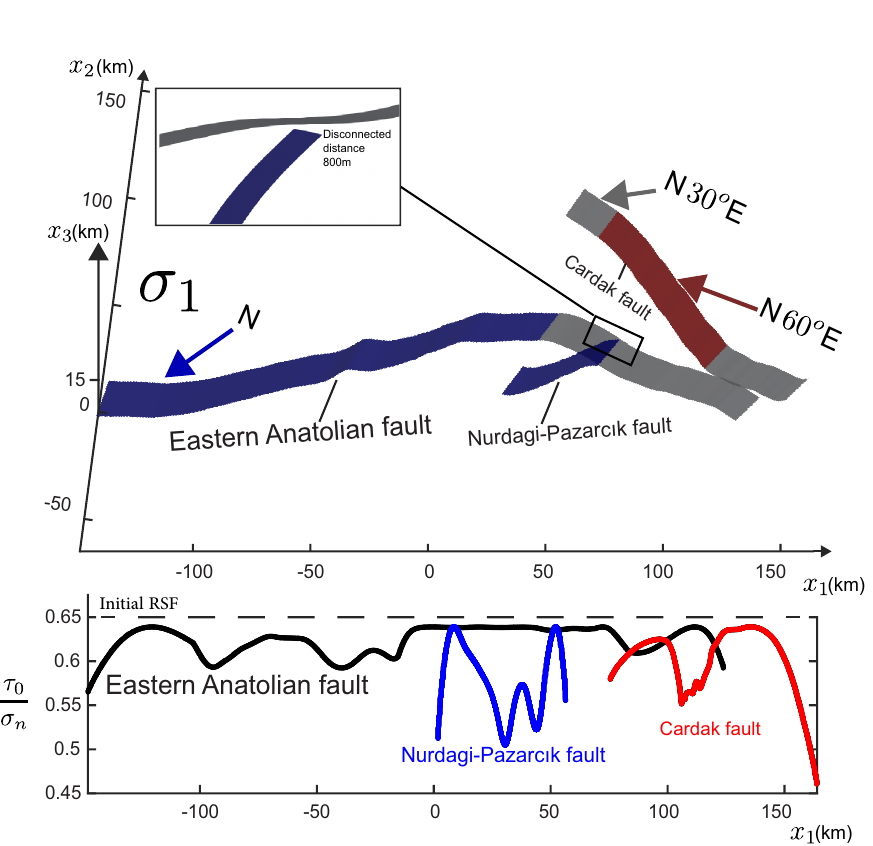}
\caption{Geometry of the fault system involved in the 2023 Kahramanmaraş–Türkiye earthquakes and stress field setting in this study. The bottom panel shows the initial ratio of shear to normal traction on each fault.}
\label{fig:turkey_geometry}
\end{figure}

We construct our fault model (Fig. \ref{fig:turkey_geometry}) according to the displacement discontinuity defined in the InSAR image (Raimbault, Jolivet and Aochi, personal communication, 2024), with a fixed dip of $90^\circ$ down to 15 km. The NPF and EAF are separated by a distance of 800 meters and are not connected. The frame is rotated clockwise by $30^\circ$, and the new Cartesian coordinates (X, Y) used in the simulations roughly correspond the fault parallel and normal directions \citep{Aochi2024}. Principal horizontal stress orientations are setting segmented, as shown in Fig. \ref{fig:turkey_geometry}. We also smoothed the stress orientations in order to have a reasonable range of shear and normal traction. Friction, material properties and principal stress values and stress rates are shown in Table 2. A stress field is essential for defining the initial and loading conditions. We adopt a segmented principal stress tensor (depicted in different colors in Fig. \ref{fig:turkey_geometry}), keeping it as simple as possible while referencing the local strain-rate tensor from \citet{weiss2020}.

\begin{figure}
\centering
\includegraphics[width=\textwidth]{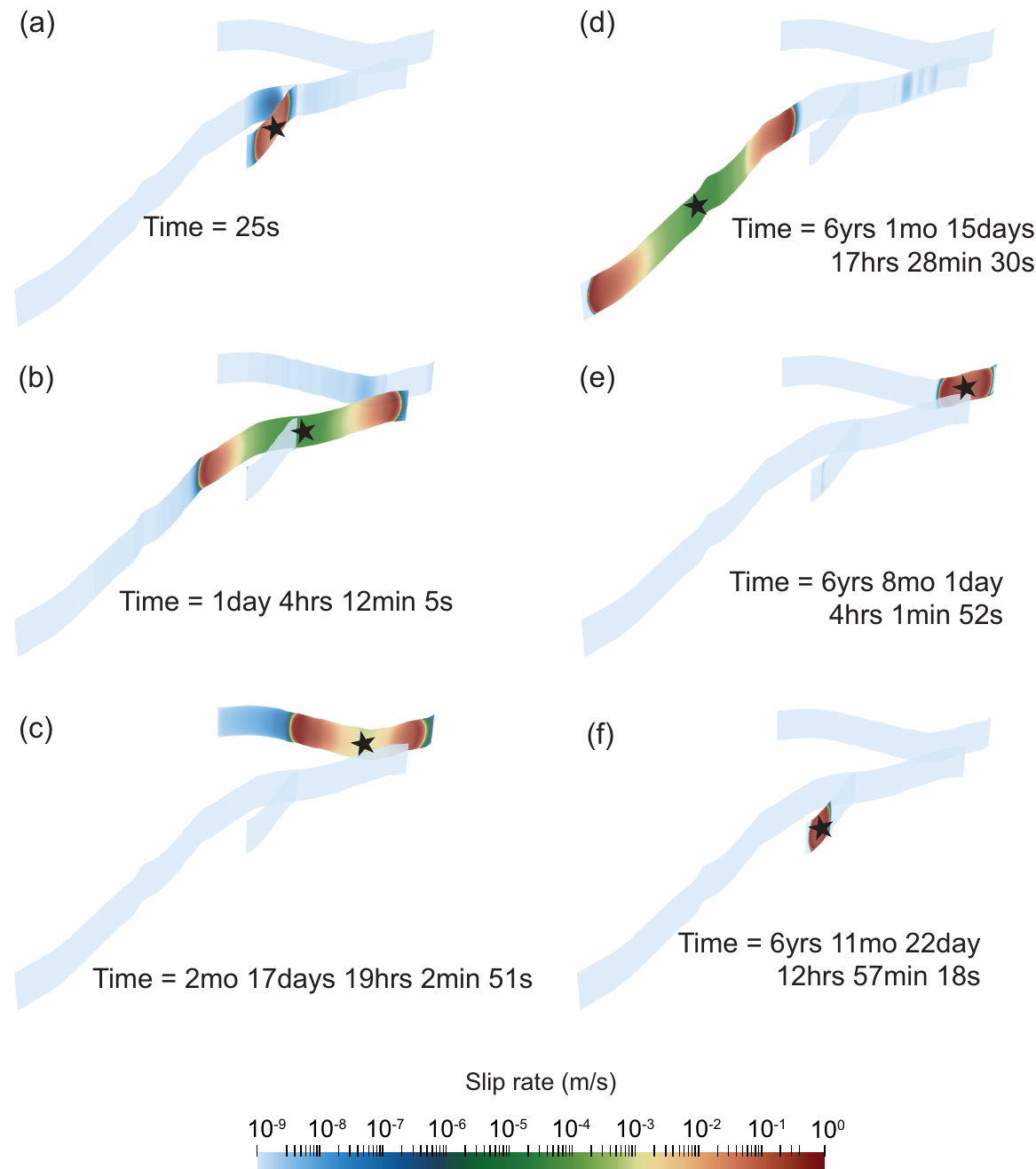}
\caption{The slip rate snapshots in Turkish model for the first and second earthquake sequences, along with the corresponding times. (a)(b)(c) The first sequences rupture in order of NPF, EAF and Cardak fault (d)(e)(f) The second sequences rupture in order of EAF, Çardak fault and NPF. The stars denote the hypocenters.}
\label{fig:turkey_result1}
\end{figure}

The process zone length \(L_b\) is 2400 m according to the model parameters we chose. We use a grid size of 400 m, corresponding to \(L_b/6\), ensuring sufficient resolution for the simulation. The model consists of 105,722 triangular elements. To simulate the 2023 Turkey doublet earthquakes, 6500 time steps are required, and the computation is completed within 2 hours. The rupture process and timing are illustrated in Fig. \ref{fig:turkey_result1} (a) (b) (c). The earthquake initially nucleated on the Nurdağı-Pazarcık fault, then propagated bilaterally along the Eastern Anatolian Fault, eventually triggering rupture on the Çardak fault. We don't aim to reproduce exact timing, due to the use of quasi-dynamic simulation and the uncertainty in the loading rate. Instead, we focus on fault interactions within a multi-fault system, demonstrating that even with spatially uniform friction properties, we can reproduce the complex slip sequences. We specifically note that delayed triggering, overall moment and average slip distribution and rupture complexity here are purely the consequence of complex fault geometry.
 
\begin{table}\label{tab:Model parameters for turkey}

 \caption{Model and numerical parameters for the 2023 Kahramanmaraş – Türkiye Earthquakes}
 \centering

 \begin{tabular}{c c c}
 \hline
  Parameters & Description & Value \\
 \hline
   $a$ & direct effect parameter in RSF & 0.003   \\
   $b$ & evolution effect parameter in RSF & 0.01  \\
   $D_{c}$ & characteristic slip  & 0.04 m  \\
   $V_{ref}$ &  reference velocity for RSF &  $10^{-6}$ m/s  \\
   $f_{0}$ & reference friction coefficient for RSF  & 0.6    \\
   $\mu$ & shear modulus & $3 \times 10^{10}$ Pa\\
    $\rho$ & density & 2670 kg/$m^{3}$\\
    $C_{s}$ & shear wave velocity &  3464 m/s\\
    $V_{0}$& initial slip rate &  $10^{-9}$ m/s\\
    $\sigma_{1}$ & maximum principal stress &  100 MPa\\
    $\sigma_{2}$ & vertical principal stress &  50 MPa\\
    $\sigma_{3}$ & minimum principal stress &  30 MPa\\
    $\dot{\sigma_{1}}$ & maximum principal stress rate &  0.2 Pa/s\\
    $\dot{\sigma_{2}}$ & vertical principal stress rate &  0.05 Pa/s\\
    $\dot{\sigma_{3}}$ & minimum principal stress rate&  0.03 Pa/s\\
 \hline
% \multicolumn{2}{l}{$^{a}$Footnote text here.}
 \end{tabular}
 \end{table}

\begin{figure}
\centering
\includegraphics[width=\textwidth]{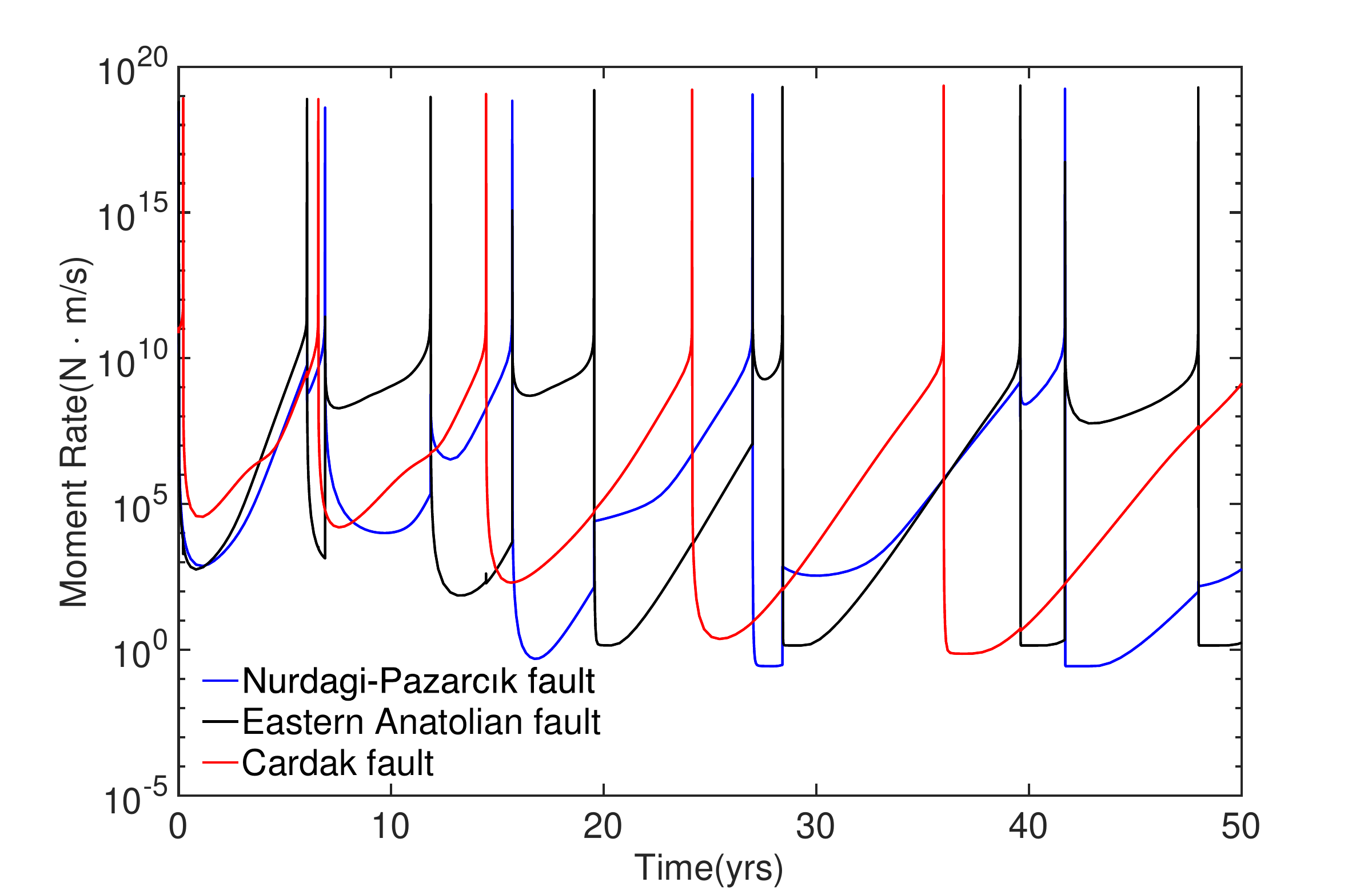}
\caption{Moment rate evolution on each fault segment over 50 years.}
\label{fig:turkey_cycle}
\end{figure}

We then simulate multiple earthquake cycles on this fault system, solving for six earthquakes on the main faults, as shown in Fig. \ref{fig:turkey_cycle}. The simulation runs for 68,000 time steps and completes within one day. We observe two distinct rupture sequences: the first is depicted in Fig. \ref{fig:turkey_result1} (a), (b), and (c), while the second is shown in panels (d), (e), and (f). Our results show the earthquakes on EAF are more frequent than those on Cardak fault and NPF. Due to the interaction between three faults, we have complex earthquake sequences in the following earthquake cycles.

\newpage
\section{Discussion and Conclusion}

In this paper, we introduce a quasi-dynamic earthquake cycle model of 3D geometrical complex fault systems that is governed by rate and state friction law. Using a boundary element method accelerated by Hierarchical Matrices, we convert force equilibrium on faults into a series of ordinary differential equations (ODEs) to produce the evolution of slip rate, rake angle, shear and normal tractions, and moment rate on the faults. We carefully benchmarked our model with a static penny-shaped crack under constant loading and also the same with an elastic crack surrounded by a cohesive zone (Dugdale crack). We also cross-validated with the SCEC SEAS BP4-QD benchmark to validate the code accuracy during earthquake cycles. For an effective simulation, having a high-quality mesh is crucial. We also provide insights on selecting numerical parameters, like H-mat properties and tolerance for ODE solvers, to strike a balance between manageable computing time and accurate results.

Unlike conventional modeling approaches that focus on single planar faults, this model allows us to simulate earthquake behavior in complex fault systems with multiple faults and nonplanar geometries. Our goal is to clarify how complex fault geometry influences earthquake sequences. These geometrical complexities can be accurately defined using geological and geophysical methods, enabling us to simulate real-world fault systems and apply a physics-based model to estimate earthquake magnitudes, locations, and frequencies. We applied this approach to the 2023 earthquakes in Turkey, considering the interactions between three faults. With our method, we analyzed six earthquakes in just one day—compared to 24 days with traditional techniques that don't use H-Matrix acceleration. To capture timing accurately and study supershear ruptures, a fully dynamic model is needed. Notably, these complex sequences emerge from spatially uniform friction properties, highlighting the importance of considering geometric complexities. Future research should focus on combining both frictional heterogeneity and geometric complexity in earthquake cycle simulations to better understand realistic earthquake sequences.

Our current model focuses only on the full space kernel and doesn't consider the effect of the free surface. \citet{nikkhoo2015} have developed a solution for the half space kernel with triangular dislocation. We plan to improve our model by including this half space approach and also considering the impact of topography.

\section{Acknowledgements}
JC, MA and HSB gratefully acknowledge the European Research Council (ERC) for its full support of this work through the PERSISMO grant (No. 865411). The numerical simulations presented in this study were performed on the MADARIAGA cluster, also supported by the ERC PERSISMO grant. We thank Pierpaolo Dubernet for his assistance with the use and management of the cluster. BL gratefully acknowledges funding to the EMOD project (Engineering model for hydraulic stimulation) which benefits from a grant (research contract no. SI/502081-01) and an exploration subsidy (contract no. MF-021-GEO-ERK) of the Swiss federal office of energy for the EGS geothermal project in Haute-Sorne, canton of Jura.

\section{Data Availability}
All the simulation results and codes related to figures in this paper are available in \url{https://zenodo.org/doi/10.5281/zenodo.14800648}.

\clearpage
\setcounter{section}{0}
\renewcommand\thesection{Appendix \Alph{section}}
\renewcommand\thesubsection{\thesection.\arabic{subsection}}
\renewcommand{\theequation}{A.\arabic{equation}}
\setcounter{equation}{0}

\begin{center}
\LARGE {\bf Appendix}
\end{center}

\section{ODE system} \label{ode}

We differentiate the force equilibrium equation in Eq. \ref{tf},
\begin{equation}
\dot \tau^f = -\dot f \cdot (T_n-p)-f\cdot (\dot T_n-\dot p),
\end{equation}
where $T_{n}$  is normal traction and $p$ is pore pressure.

We differentiate the rate and state friction in Eq. \ref{friction}
\begin{equation}
%\dot f = \dfrac{a}{V}\dot V+\dfrac{b}{\theta}\dot\theta\\
\dot f = \dfrac{\partial{f}}{\partial{V}}\dot V+ \dfrac{\partial{f}}{\partial{\theta}}\dot \theta.
\end{equation}

Let

\begin{equation}
\dfrac{\partial{f}}{\partial{V}} = A
\end{equation}
\begin{equation}
\dfrac{\partial{f}}{\partial{\theta}} = B.
\end{equation}

It is important to note that our ODE solver operates in a local coordinate system, so we transform the coordinates accordingly. We differentiate Eq. \ref{t1}, \ref{t2}, \ref{t3}, \ref{tf}, \ref{theta} and get a set of ODEs: 
\begin{gather*}
\dot\tau \cos{\alpha} - \dot\alpha \tau \sin{\alpha} = \dot\tau_1^{el}+\dot\tau_1^{load}+\eta_s (\dot V \cos{\alpha}-\dot \alpha V \sin{\alpha})\\
\dot\tau \sin{\alpha} + \dot\alpha \tau \cos{\alpha}= \dot\tau_2^{el}+\dot\tau_2^{load}+\eta_s (\dot V \sin{\alpha}+\dot \alpha V \cos{\alpha})\\
\dot T_n = \dot \tau_3^{el}+\dot \tau_3^{load}\\
\dot \tau = -(A \dot V+B \dot\theta) \cdot (T_n-p)-f \cdot (\dot T_n-\dot p)\\
\dot \theta= 1-\dfrac{V\theta}{D_c},
\end{gather*}
where, $\tau_1$ and $\tau_2$ are tractions along $\mathbf{e_1}$ and $\mathbf{e_2}$ in local coordinate system, $\alpha$ is the angle between $\mathbf{e_1}$ and the slip vector. These ODEs can be written in matrix-vector multiplication form as follows:

\begin{equation}
\begin{bmatrix}
-\eta_s \cos{\alpha} & \eta_s V\sin{\alpha}-\tau \sin{\alpha} & 0 &\cos{\alpha}&0\\
-\eta_s \sin{\alpha}&-\eta_s V\cos{\alpha}+\tau \cos{\alpha} &0&\sin{\alpha}&0\\
0&0&1&0&0\\
A(T_n-p)&0&f&1&B(T_n-p)\\
0&0&0&0&1
\end{bmatrix}
\dfrac{\partial}{\partial t}
\begin{bmatrix}
V\\
\alpha\\
T_n\\
\tau\\
\theta
\end{bmatrix}
=\begin{bmatrix}
\dot\tau_1^{el}+\dot\tau_1^{load}\\
 \dot\tau_2^{el}+\dot\tau_2^{load}\\
 \dot\tau_3^{el}+\dot\tau_3^{load}\\
f\dot p\\
\dfrac{d\theta}{dt}
%1-\dfrac{V\theta}{D_c}
\end{bmatrix}
\end{equation}

We can write this ODE system explicitly
\begin{equation}
\dfrac{\partial}{\partial t}
\begin{bmatrix}
V\\
\alpha\\
T_n\\
\tau\\
\theta
\end{bmatrix}
=
\begin{bmatrix}
\dfrac{\cos{\alpha}}{D_1} & \dfrac{\sin{\alpha}}{D_1} & \dfrac{f}{D_1} & -\dfrac{1}{D_1} & \dfrac{B(T_n-p)} {D_1}\\
\dfrac{\sin{\alpha}}{\eta V - \tau} & -\dfrac{\cos{\alpha}}{\eta V- \tau} & 0 & 0 & 0\\
0 & 0 & 1 & 0 &0\\
\dfrac{A\cos{\alpha}(p-T_n)}{D_1} & \dfrac{A\sin{\alpha}(p-T_n)}{D_1}  & \dfrac{f \eta}{D_1}& -\dfrac{\eta}{D_1} & \dfrac{B\eta (T_n-p)}{D_1}\\
0&0&0&0&1
\end{bmatrix}
\begin{bmatrix}
\dot\tau_1^{el}+\dot\tau_1^{load}\\
 \dot\tau_2^{el}+\dot\tau_2^{load}\\
 \dot\tau_3^{el}+\dot\tau_3^{load}\\
f\dot p\\
\dfrac{d\theta}{dt}
\end{bmatrix}
\end{equation}

where $D_{1}=A(p-T_n)-\eta$.

For non-regularied rate and state friction law, friction is
\begin{equation}
f = f_0+a \log\left(\dfrac{V}{V_{ref}}\right)+b\log\left(\dfrac{\theta V_{ref}}{D_c}\right)
\end{equation}
\begin{gather*}
A=\dfrac{\partial{f}}{\partial{V}} =\dfrac{a}{V}\\
B=\dfrac{\partial{f}}{\partial{\theta}}= \dfrac{b}{\theta}.\\
\end{gather*}

For regularized form
\begin{equation}
f = a~\arcsinh{\left\{\dfrac{V}{2V_{ref}}\exp{\left[\dfrac{f_0}{a}+\dfrac{b}{a}\log{\left(\dfrac{\theta V_{ref}}{D_c}\right)}\right]}\right\}}
\end{equation}

\begin{equation}
z = \dfrac{1}{2V_{ref}}\exp{\left[\dfrac{f_0}{a}+\dfrac{b}{a}\log{\left(\dfrac{\theta V_{ref}}{D_c}\right)}\right]}
\end{equation}

\begin{equation}
A=\dfrac{\partial{f}}{\partial{V}} = \dfrac{a}{\sqrt{z^{-2}+V^2}} 
\end{equation}
\begin{equation}
B=\dfrac{\partial{f}}{\partial{\theta}} =\dfrac{\partial f}{\partial V}\cdot \dfrac{bV}{a\theta}.
\end{equation}
\section{coordinate transformation}

In the fault-based coordinate system, the fault normal vector \(\mathbf{n}\) is defined as perpendicular to the fault plane. Once the fault geometry is specified, \(\mathbf{n}\) can be readily calculated. 

\begin{equation}
\mathbf{n} = \mathbf{e_3} = (e_{31},e_{32},e_{33}).
\end{equation}

Strike $\mathbf{s}$ is defined as the line of intersection between a horizontal plane and the fault surface, 
\begin{equation}
\mathbf{s} = \dfrac{\mathbf{x_3} \times \mathbf{n}}{||\mathbf{x_3} \times \mathbf{n}||} = \left |\begin{array}{ccc}
i &j   & k \\
0 & 0 &1 \\
e_{31} &e_{32} &e_{33}  \\
\end{array}\right| / ||\mathbf{x_3} \times \mathbf{n}||= \dfrac{1}{\sqrt{e_{31}^2+e_{32}^2}}\left ( -e_{32},e_{31},0\right).
\end{equation}
We define $\mathbf{s}$ is positive for left-lateral slip.

Dip is the angle between horizontal plane and fault surface. We define $\mathbf{d}$ to point upward. Since $(\mathbf{s},\mathbf{n},\mathbf{d})$ is a Cartesian coordinate system, 
\begin{equation}
\mathbf{d} = \mathbf{n} \times \mathbf{s}= \dfrac{1}{\sqrt{e_{31}^2+e_{32}^2}} \left |\begin{array}{ccc}
i & j &k \\
e_{32}&-e_{31}&0 \\
e_{31} &e_{32} &e_{33}  \\
\end{array}\right|= \dfrac{1}{\sqrt{e_{31}^2+e_{32}^2}}\left (-e_{31}e_{33},-e_{32}e_{33},e_{31}^2+e_{32}^2\right).
\end{equation}
This is written in the global system.

Slip rate vector $\vec{V}$ in local system is ($V_{e_1},V_{e_2},{V_{e_3}}$), in fault-based system is $(V_{s},V_{d},{V_n})$. In the global system, the slip vector $\vec{V}$ can be written with the basis $(\mathbf{e_1},\mathbf{e_2},\mathbf{e_3})$, $\vec{V} = \vec{V_{local}} = V_{e_1}\mathbf{e_1}+V_{e_2}\mathbf{e_2}+ V_{e_3}\mathbf{e_3}$. With basis $(\mathbf{s},\mathbf{d},\mathbf{n})$, $\vec{V} = \vec{V_{fault}} = V_{s}\mathbf{s} +  V_{d}\mathbf{d}{+V_{n}\mathbf{n}}$.

In the strike direction, considering the vector algebra relation $A \cdot (B \times C) = C \cdot (A \times B)$, we can write:
\begin{equation}
V_s = \vec{V_{local}} \cdot \mathbf{s} = \vec{V_{local}} \cdot \dfrac{\mathbf{x_3} \times \mathbf{n}}{||\mathbf{x_3} \times \mathbf{n}||} = \dfrac{1}{\sqrt{e_{31}^2+e_{32}^2}} \mathbf{x_3} \cdot (\mathbf{n}\times \vec{V_{local}}).
\end{equation}

Because $\mathbf{e_1}, \mathbf{e_2}, \mathbf{e_3}$ are orthognal, we can write $ \mathbf{e_3}\times \mathbf{e_1} = \mathbf{e_2}$, $\mathbf{e_3} \times \mathbf{e_2}=-\mathbf{e_1}$

\begin{equation}
 \mathbf{n}  \times \vec{V_{local}} = \mathbf{e_3}\times (V_{e_1}\mathbf{e_1}+V_{e_2}\mathbf{e_2}{+ V_{e_3}\mathbf{e_3}}{}) = V_{e_1}\mathbf{e_2}-V_{e_2}\mathbf{e_1}.
\end{equation}

Then, 
%\colorbox{red}{
\begin{equation}
V_s = \dfrac{1}{\sqrt{e_{31}^2+e_{32}^2}} \mathbf{x_3} \cdot (V_{e_1}\mathbf{e_2}-V_{e_2}\mathbf{e_1}) =  \dfrac{1}{\sqrt{e_{31}^2+e_{32}^2}}(V_{e_1}e_{23}-V_{e_2}e_{13})\label{transformation1}.
\end{equation}
%}

In the dip direction, we can write:
\begin{equation} 
V_d = \vec{V_{local}} \cdot \mathbf{d} = \vec{V_{local}} \cdot (\mathbf{n} \times \mathbf{s})= \mathbf{s} \cdot (\vec{V_{local}}\times\mathbf{n} ) = \dfrac{1}{\sqrt{e_{31}^2+e_{32}^2}} (\mathbf{x_3}\times \mathbf{n}  ) \cdot ( \vec{V_{local}}\times \mathbf{n}).\label{transformation2}
\end{equation}

Considering the vector algebra relation $(A\times B) \cdot (C \times D) = (A \cdot C)(B \cdot D)-(B \cdot C)(A \cdot D) $. Slip rate vector has no opening component, so $\mathbf{n} \cdot \vec{V_{local}} = 0$ 

Then,
%\colorbox{red}{
\begin{equation}
V_d = \dfrac{1}{\sqrt{e_{31}^2+e_{32}^2}} \mathbf{x_3} \cdot \vec{V_{local}} = \dfrac{1}{\sqrt{e_{31}^2+e_{32}^2}} (V_{e_1}e_{13}+V_{e_2}e_{23}{+V_{e_3}e_{33}}).
\end{equation}
%}

We can write the slip rate in matrix format as follows:
\begin{equation}
\left [\begin{array}{c}
V_s \\
V_d  \\
{V_n} \\
\end{array}\right]= \left [\begin{array}{ccc}
\dfrac{e_{23}}{\sqrt{e_{31}^2+e_{32}^2}} & \dfrac{-e_{13}}{\sqrt{e_{31}^2+e_{32}^2}} &0 \\
\dfrac{e_{13}}{\sqrt{e_{31}^2+e_{32}^2}} &\dfrac{e_{23}}
{\sqrt{e_{31}^2+e_{32}^2}} &0  \\
0&0&1 \\
\end{array}\right]\left [\begin{array}{c}
V_{e_1} \\
V_{e_2}\\
V_{e_3} \\
\end{array}\right].
\end{equation}

The matrix transformation from local system to fault-based system R is 
\begin{equation}
R = \left [\begin{array}{ccc}
\dfrac{e_{23}}{\sqrt{e_{31}^2+e_{32}^2}} & \dfrac{-e_{13}}{\sqrt{e_{31}^2+e_{32}^2}} &0 \\
\dfrac{e_{13}}{\sqrt{e_{31}^2+e_{32}^2}} &\dfrac{e_{23}}{\sqrt{e_{31}^2+e_{32}^2}} &0  \\
0&0&1 \\
\end{array}\right].
\end{equation}

The inverse of R can help us transform from the fault-based system to the local system:
\begin{equation}
R^{-1}= \left [\begin{array}{ccc}
\dfrac{e_{23}\sqrt{e_{31}^2+e_{32}^2}}{e_{23}^2+e_{13}^2}  &\dfrac{e_{13}\sqrt{e_{31}^2+e_{32}^2}}{e_{23}^2+e_{13}^2}& 0 \\
\dfrac{-e_{13}\sqrt{e_{31}^2+e_{32}^2}}{e_{23}^2+e_{13}^2}& \dfrac{e_{23}\sqrt{e_{31}^2+e_{32}^2}}{e_{23}^2+e_{13}^2}& 0\\
0&0&1 \\
\end{array}\right].
\end{equation}

Thus, the slip rate can be written in the local coordinate system as follows:
\begin{equation}
\left [\begin{array}{c}
V_{e_1} \\
V_{e_2}\\
V_{e_3} \\
\end{array}\right] =
\left [\begin{array}{ccc}
\dfrac{e_{23}\sqrt{e_{31}^2+e_{32}^2}}{e_{23}^2+e_{13}^2}  &\dfrac{e_{13}\sqrt{e_{31}^2+e_{32}^2}}{e_{23}^2+e_{13}^2} & 0\\
\dfrac{-e_{13}\sqrt{e_{31}^2+e_{32}^2}}{e_{23}^2+e_{13}^2}& \dfrac{e_{23}\sqrt{e_{31}^2+e_{32}^2}}{e_{23}^2+e_{13}^2}&0\\
0&0&1 \\
\end{array}\right]
\left [\begin{array}{c}
V_s \\
V_d  \\
{V_n} \\
\end{array}\right].
\end{equation}

Therefore,
%\colorbox{red}{
\begin{equation}
V_{e_1} = \dfrac{\sqrt{e_{31}^2+e_{32}^2}}{e_{23}^2+e_{13}^2} (V_s e_{23}+V_d e_{13})\label{transformation3}
\end{equation}
%}
%\colorbox{red}{
\begin{equation}
V_{e_2} = \dfrac{\sqrt{e_{31}^2+e_{32}^2}}{e_{23}^2+e_{13}^2} (-V_s e_{13}+V_d e_{23})\label{transformation4}
\end{equation}
%}

\begin{equation}
\alpha = \tan^{-1}{\left(V_{e_2}/V_{e_1}\right)};
\lambda = \tan^{-1}{\left(V_{d}/V_{s}\right)}.
\end{equation}

\clearpage 
\section{Penny-shaped crack benchmark}
\label{penny_shaped}

\begin{figure}[h!]
\centering
\includegraphics[width=0.9\textwidth]{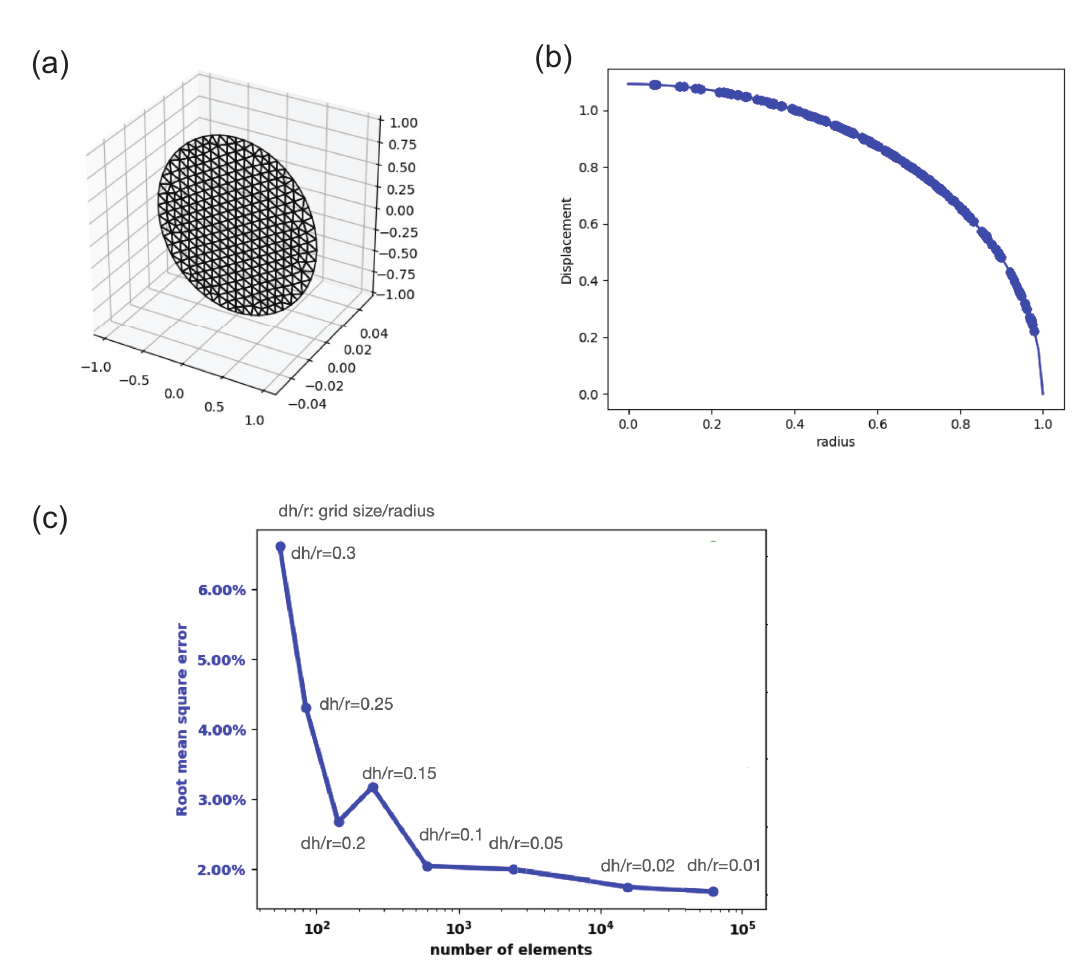}
\caption{(a) A penny-shaped crack with a radius of 1, loaded by far-field shear stress and discretized with triangular elements. (b) Displacement analytical solution from \citet{Segedin1951} along the crack radius shown in (a), with points representing the displacement at each discretized triangle. (c) The root mean square error between the calculated traction and analytical traction as the number of elements increases.}
\label{fig:static_benchmark_penny_shape}
\end{figure}
\label{lastpage}

\end{document}